\documentclass{osa-article}

\journal{boe}

\articletype{Research Article}

\usepackage{xcolor,subcaption,wrapfig,booktabs}
\usepackage[export]{adjustbox}
\usepackage[shortlabels]{enumitem}
\usepackage{centernot}
\usepackage{mathtools}
\usepackage{stmaryrd}
\usepackage{appendix}
\usepackage{xfrac}

\usepackage{scalerel}
\usepackage{tikz}
\usetikzlibrary{svg.path}

\definecolor{orcidlogocol}{HTML}{A6CE39}
\tikzset{
  orcidlogo/.pic={
    \fill[orcidlogocol] svg{M256,128c0,70.7-57.3,128-128,128C57.3,256,0,198.7,0,128C0,57.3,57.3,0,128,0C198.7,0,256,57.3,256,128z};
    \fill[white] svg{M86.3,186.2H70.9V79.1h15.4v48.4V186.2z}
                 svg{M108.9,79.1h41.6c39.6,0,57,28.3,57,53.6c0,27.5-21.5,53.6-56.8,53.6h-41.8V79.1z M124.3,172.4h24.5c34.9,0,42.9-26.5,42.9-39.7c0-21.5-13.7-39.7-43.7-39.7h-23.7V172.4z}
                 svg{M88.7,56.8c0,5.5-4.5,10.1-10.1,10.1c-5.6,0-10.1-4.6-10.1-10.1c0-5.6,4.5-10.1,10.1-10.1C84.2,46.7,88.7,51.3,88.7,56.8z};
  }
}

\newcommand\orcidicon[1]{\href{https://orcid.org/#1}{\mbox{\scalerel*{
\begin{tikzpicture}[yscale=-1,transform shape]
\pic{orcidlogo};
\end{tikzpicture}
}{|}}}}

\usepackage{hyperref} 

\renewcommand{\arraystretch}{1.8} 
\newcommand\etal{\emph{et al.}}

\makeatletter
\newcommand{\xMapsto}[2][]{\ext@arrow 0599{\Mapstofill@}{#1}{#2}}
\def\Mapstofill@{\arrowfill@{\Mapstochar\Relbar}\Relbar\Rightarrow}
\makeatother

\newcommand\eg{\emph{e.g.}}
\newcommand\ie{\emph{i.e.}}

\begin{document}

    


\title{Learning to Reconstruct Confocal Microscopy Stacks from Single Light Field Images}

\author{Josue Page\authormark{1,4,*}\orcidicon{0000-0002-5848-3063}, Federico Saltarin\authormark{2}\orcidicon{0000-0002-4320-931X}, Yury Belyaev\authormark{3} \orcidicon{0000-0001-8318-2951}, Ruth Lyck\authormark{2} \orcidicon{0000-0002-6479-4837} and Paolo Favaro\authormark{1} \orcidicon{0000-0003-3546-8247}}

\address{\authormark{1}Department of Computer Science, University of Bern, Switzerland\\
\authormark{2}Theodor Kocher Institute, University of Bern, Switzerland\\
\authormark{3}Microscopy Imaging Center, University of Bern, Switzerland\\
\authormark{4}Department of Informatics, Technical University of Munich, Germany}

\email{\authormark{*}josue.page@inf.unibe.ch} 



\begin{abstract}
We present a novel deep learning approach to reconstruct confocal microscopy stacks from single light field images. To perform the reconstruction, we introduce the LFMNet, a novel neural network architecture inspired by the U-Net design \cite{ref:unet}. It is able to reconstruct with high-accuracy a $112 \times 112 \times 57.6 \mu m^3$ volume ($1287\times 1287\times 64$ voxels) in $50ms$ given a single light field image of $1287\times 1287$  pixels, thus dramatically reducing $720$-fold the time for confocal scanning of assays at the same volumetric resolution and $64$-fold the required storage. 
To prove the applicability in life sciences, our approach is evaluated both quantitatively and qualitatively on mouse brain slices with fluorescently labelled blood vessels. Because of the drastic reduction in scan time and storage space, our setup and method are directly applicable to real-time \textit{in vivo} 3D microscopy. We provide analysis of the optical design, of the network architecture and of our training procedure to optimally reconstruct volumes for a given target depth range. To train our network, we built a data set of $362$ light field images of mouse brain blood vessels and the corresponding aligned set of 3D confocal scans, which we use as ground truth. The data set will be made available for research purposes \cite{ref:MiceLFMDataset}. 
\end{abstract}


\section{Introduction}

Confocal microscopes can provide high-quality scans of volumes, which find extensive application in life sciences. However, confocal imaging requires high excitation power, as most of the fluorescence is blocked by a pinhole, which results in high photo-toxicity and photo-bleaching. It is also very time-consuming due to the voxel-wise acquisition of the image, and requires a substantial amount of storage for the acquired data. Moreover, it is unsuitable for fast \textit{in vivo} imaging due to its spatio-temporal distortion of moving samples. For instance, imaging of processes, such as calcium signaling during neuronal conduction or the beating of a zebrafish heart, is often affected by reconstruction artifacts \cite{ref:Mariani2019}. 

We propose to use light field microscopy  \cite{ref:Levoy2006} in combination with a deep learning approach to address the above shortcomings and to ensure high quality reconstructions of volumes. Light field microscopy is a technique that turns any wide field microscope into a scan-less single shot 3D microscope. This is achieved by placing a micro-lens array (MLA) in the optical path and by using an algorithm (\eg deconvolution) to reconstruct the observed volume from the light field (LF) image. 
It enables advanced computational applications, such as \textit{in vivo} calcium imaging of neural activity, in immobilized \cite{ref:Grosenick2009,ref:Grosenick2017,ref:Nobauer2017,ref:Pegard2016a,ref:Prevedel2014,ref:CruzPerez2015,ref:Zhang2020} and freely moving animals \cite{ref:Skocek2018,ref:LinCong,ref:Aimon2019,ref:Shaw2018}. Also, it has been used to image mouse brain tissue, which has high scattering up to $380\mu m$ in depth   \cite{ref:Nobauer2017}.

Light field microscopes (LFM) trade off spatial information (the image resolution) for depth information. This partly explains why current super-resolution algorithms for the reconstruction of volumes from a single light field image are not able to match the resolution of the confocal scan data in all three spatial axes \cite{ref:Levoy2006}. Moreover, these reconstruction algorithms are often rather slow due to their iterative nature and high computational complexity. 
Thus, the development of a method for high-resolution and fast 3D reconstruction from LFM images would provide a high-impact method in life sciences, for imaging of very fast biological processes. 

Most volume reconstruction methods are based first on devising a mathematical model of the optics and then on using such model to design a deconvolution algorithm of the LF images \cite{ref:Broxton2013,ref:Cohen2014,ref:Levoy2009,ref:Bishop2012a}. Current state of the art algorithms suffer from poor spatial resolution (mainly at the focal plane of the microscope), reconstruction artifacts and extreme computational times, which are not suitable for real-time imaging.
\begin{figure}
    \centering
    \includegraphics[width=1.07\textwidth]{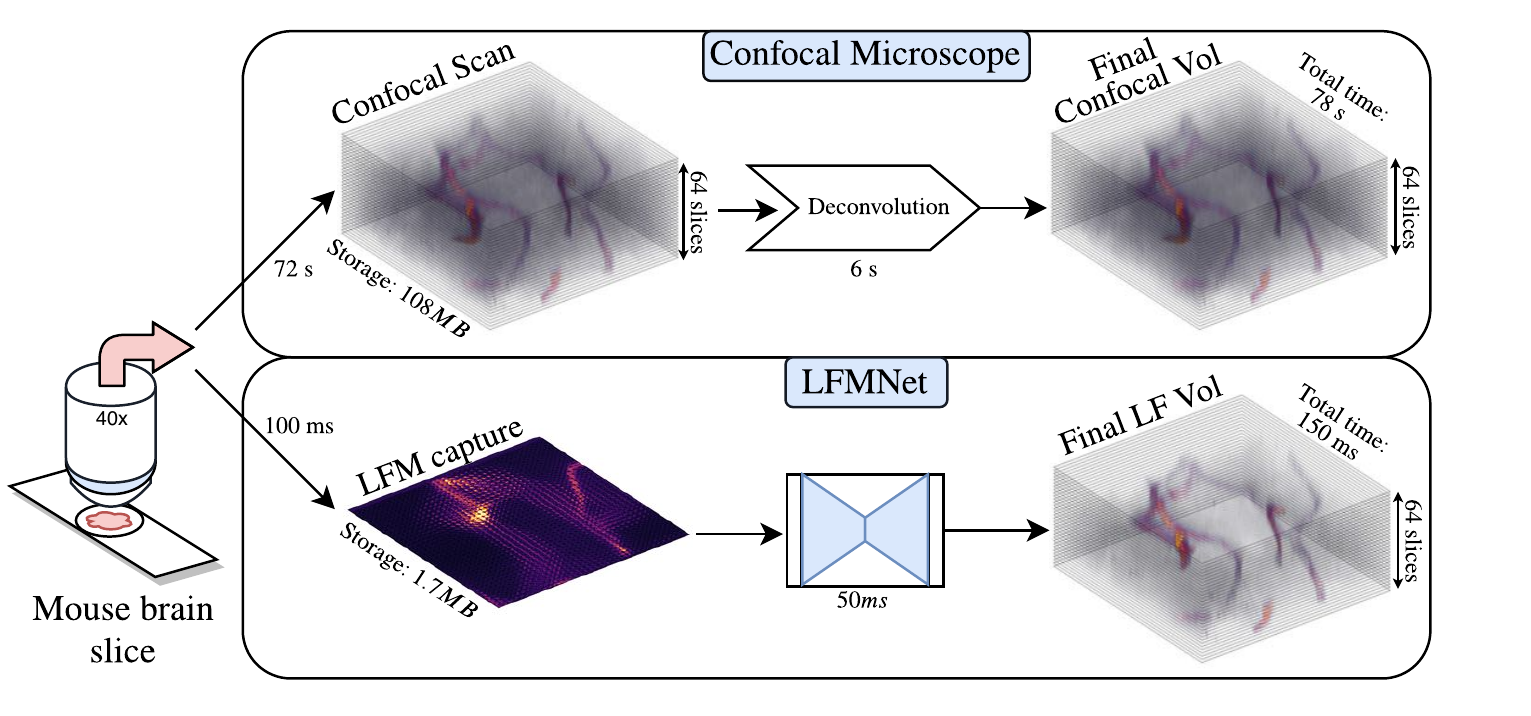}
    \caption{\textbf{Comparison of a confocal stack scan vs a scan with our LFM.} Storage, capture and computational times refer to a volume with $1287\times1287\times64$ voxels. Notice how the LFM data acquisition (bottom row) is faster both at capture time and rendering time, and requires much less storage than the confocal stack scan (top row).}
    \label{fig:syst_compare}
\end{figure}
In contrast, as shown in Fig.~\ref{fig:syst_compare}, our solution provides remarkable reconstructions of volumes at resolutions close to those obtained with confocal imaging under similar optical settings. 
Moreover, our method reduces acquisition and reconstruction times from hours to minutes, thus yielding an approach that can capture and process $10$ frames per second (mostly delayed by the camera's exposure time of $100ms$). 
We achieve this by designing LFMNet, a novel convolutional neural network (CNN) architecture, and by training it with real confocal stacks and their corresponding aligned LF images (for more details, see section~\ref{sec:DLModel}). 
Compared to previous work, where a deep learning architecture was trained on synthetic data \cite{ref:Wanga}, we use a data set of real LF images and confocal stacks. The main challenge in building and exploiting such a data set is that LF images and the corresponding confocal scan volumes must be aligned. Our alignment technique first computes the reconstruction of an approximate sharp volume via a model-based method \cite{ref:Stefanoiu:19}. Then, it matches via a correlation map the image obtained by averaging the volume along the Z-axis to the image obtained in the same way from the confocal volume. The correlation map yields a 2D shift that allows us to locate and align the LF image to the confocal scan volume. We do not need to consider further transformations (such as, \eg, in-plane rotations), because we optically calibrate the LFM sensor to that of the confocal microscope. We find experimentally that the overall accuracy of our alignment procedure is sufficient to train the LFMNet to output high quality reconstructions. We observe that the proposed approach does not require a time-consuming procedure to build the data set, as no manual labeling is required and only a small number of high-resolution images is needed ($362$ in our data set). We also illustrate the pipeline used to align the LF images to the confocal stacks in section~\ref{sec:alignment}.
The main advantage of using real data instead of synthetic data is that the trained network has the opportunity to capture a more effective image prior in the domain of interest. 
We will publicly release the data set of 362 LF images (each of $1287\times1287$ pixels, which correspond to LF spatial/angular resolutions of $33\times33\times39\times39$ elements \cite{ref:Ng2005}) with their corresponding confocal stacks ($1287\times1287\times64$ voxels) for research purposes.

A second contribution in this work is the design of the LFMNet. To better model the 4D nature of the input LF image, the first layer of our network is a 4D convolution, whose output is reshaped as an image and then fed as input to a modified U-Net architecture, where the channels are mapped to the depth axis. Moreover, the network is designed to be fully convolutional, so that it can process input LF images of different sizes, and to have a limited receptive field to avoid overfitting. These two design choices allow a computationally feasible and stable training of LFMNet on large batches of cropped regions of the LF images and the corresponding confocal stacks (for more details, see section~\ref{sec:PatchesVsfull}). 


A third contribution is the analysis of the LFM optical configuration to determine which settings yield an optimal 3D reconstruction. The placement of the MLA and of the sensor in the optical path of a LF system plays a crucial role on the amount of angular and spatial information available at the sensor, the amount of aliasing and the achievable 3D reconstruction accuracy \cite{ref:Ng2005,ref:Georgiev,ref:Aoyu2019,ref:Bishop2012a}. We carried out an extensive analysis on the possible configurations by using simulated data, and then used the Fisher information on our models of the point spread function, contrast and correlation coefficients on a USAF 1951 resolution target as performance metrics, as previously done by Broxton \etal~\cite{ref:Broxton2013} and Cohen \etal~\cite{ref:Cohen2014}. This analysis allowed us to find the best configuration for the depth range of interest (see section~\ref{sec:LFMdesign}). 



\section{Prior Work and Contributions}
Light field microscopy was first proposed by Levoy \etal~in 2006 \cite{ref:Levoy2006}. They showed the possibility of computing focal stacks and perspective views, as well as performing deconvolution of light sensitive samples. 
Later, the reconstruction quality was significantly enhanced by Broxton \etal~\cite{ref:Broxton2013} by using a wave-optics model and a super-resolution approach \cite{ref:Bishop2012a}, thus showing the limits of the angular and spatial resolution, while achieving a sampling rate 16 times the lenslet sampling rate. 
Unfortunately, when the tube-lens is focused on the MLA, all the angular information at the focal plane in object space is lost \cite{ref:Georgiev}, because the incoming light has a zero phase, and carries information only at the MLA resolution (\ie, only one voxel per micro-lens at the focal plane). In the following subsections, we discuss different approaches to mitigate this problem.

\subsubsection{LFM Designs}
In the first LFM design, the MLA was placed at the focal plane of the tube lens and the sensor at the focal plane of the MLA. This setting is called LF 1.0 and its implementation is described in \cite{ref:Ng2005}. Georgiev and Lumsdaine~\cite{ref:Georgiev,ref:Lumsdaine2009} propose instead a focused LF setting, or LF-2.0, which places the MLA and sensor relative to each other according to the thin lens equation
\begin{align}
\frac{1}{a} + \frac{1}{b} = \frac{1}{F_\text{ml}},
\label{eq:thin_lens}
\end{align}
where $F_\text{ml}$ is the focal length of the MLA, $a$ is the distance from the microscope focal plane to the MLA and $b$ from the MLA to the sensor (see Fig.~\ref{fig:MLA_Blur}). This approach yields an optical configuration, where images appear in focus at the sensor. Li \etal~\cite{ref:Aoyu2019} propose to use instead a setting such that $\frac{1}{a} + \frac{1}{b} > \frac{1}{F_\text{ml}}$, so that the induced aliasing could be exploited for 3D reconstruction purposes.  
To compensate for the loss of information at the focal plane, some approaches propose to further modify the optics.
For instance, Cohen \etal~\cite{ref:Cohen2014} and Wu \etal~\cite{ref:Wu672584} propose to introduce phase masks to shift the phase at the focal plane. Also, the works of Scrofani \etal~ \cite{ref:Scrofani2018} and Sung \cite{ref:Sung2019} propose an integral microscope without a tube-lens, that allows accessing the phase-space, and with large lenses to recover a high spatial resolution.

An alternative to solve most of the above issues, but with an extra degree of complexity, is to modify the LF acquisition step. One possibility is to employ light-sheet microscopy illumination to excite the specimen plane by plane and enhance the contrast and signal to noise ratio in the acquired images \cite{ref:Madrid-Wolff2019,ref:Wang2019}. Furthermore, Wagner \etal~\cite{ref:Wagner2018} proposed the ISO-LFM, which captures images with two synchronized LFMs positioned at 90 degrees with respect to each other. This arrangement reduces the artifact plane to a single line in the volume and allows a more isotropic resolution of the reconstruction, but still relies on individual deconvolutions for each LFM and sacrifices temporal resolution due to the light-sheet scanning.
Another approach, is proposed by Pan \etal~\cite{ref:Pan2019}, the diffraction-assisted light field microscope. They introduce a diffraction grating between the specimen and the objective, hence reducing the loss of spatial and angular information received by the sensor. Also, a mathematical model was developed and later used to recover rigid body full-field displacement measurements.
Zhang \etal~\cite{ref:Zhang2020} proposed the confocal LFM. Taking the idea from confocal microscopy, their system blocks the out of focus light coming from the specimen, achieving high axial resolution at the expense of increased exposure time. It is capable of tracking a freely moving zebra fish and capture its full brain at a rate of 6 frames per second. Their reconstructions are performed after the acquisition stage and rely on the Richardson-Lucy deconvolution algorithm, taking 60 second per volume.


\subsection{Deconvolution-Based Reconstruction}
The reconstruction of a volume from a light field image can be cast as a deconvolution problem once the point spread function (PSF) of the optics is given or estimated. The main challenge in solving deconvolution problems is that they are very sensitive to small errors in the data (a LF image in this case) and have ambiguities in the solution space. To resolve these issues, it is common to use a regularized approach, where information about the solution space is used to better constrain the problem and favor only plausible solutions. 
Multiple works have explored the PSF modeling and deconvolution in the Fourier domain \cite{ref:Liu:19,ref:Uo2019,ref:Guo:19}, taking advantage of the Fourier properties of wave optics. Lu \etal~\cite{ref:Lu2019} developed a phase space
method, which deconvolves a LF image by converting it to up-sampled views (this data arrangement is further discussed in Appendix~\ref{sec:A:dimensionality}). 
This method reconstructs a volume without the zero plane artifacts and enables faster convergence against the LF deconvolution of Broxton \etal~\cite{ref:Broxton2013}. Nevertheless, the computational time of their algorithm is still unsuitable for real-time applications. This team also propose the artifact-free deconvolution method \cite{Ref:Lu_artifact_free}. 
Stefanoiu \etal~\cite{ref:Stefanoiu:19} implemented an aliasing-aware deconvolution method that adds an additional filtering step to \cite{ref:Broxton2013} at every iteration of the deconvolution, which removes artifacts, but still incurs a high computational time.

Even though these methods have enhanced the quality of the reconstructions, they still depend on an accurate LF PSF computation, which is difficult to achieve. Furthermore, capturing the aberrations, misalignment and exact parameters of the microscope within the PSF is extremely challenging and hard to validate. Finally, deconvolution is very memory and time demanding due to the high dimensionality and complexity of the LF data.

\subsection{Deep Learning-Based Reconstruction}
Deep learning approaches applied to LFMs have the capability of learning very advanced data priors from the microscope and from the observed geometry, without requiring an explicit mathematical model of the LFM optics, of the light scattering properties of the 3D volume and of the illumination conditions. 
Instead of providing a hand-crafted approximation of such a model, deep learning approaches capture the data prior by training a general purpose neural network with a large data set of input-output examples. The challenges of this approach then lie in how the network is designed and trained.

Because the LF image is a 2D rearrangement of a 4D function (2 dimensions for the spatial coordinates and 2 for the angular coordinates), it is useful to consider its structure when designing the neural network.
Based on this principle, Wang \etal~\cite{ref:Wanga} proposed a network using a View-to-Channel (V2C) transformation, where the LF image is reshaped into a 1D list of views (see section~\ref{sec:A:dimensionality}). 
One of the main advantages of this approach is that the V2C transformation preserves a direct connection between the original structure of the input data and that of the output, and thus the complexity of the reconstruction task for the network is relatively small. 
However, this approach only works with a fixed LF image size. Moreover, the 1D mapping of the angular domain destroys the original angular lattice and this limits the learning capabilities of the network in this domain.
Also, their network was trained on simulated light fields (with the model of Broxton \etal \cite{ref:Broxton2013}) from confocal stacks, limiting the possible prior LFM information available to the network, such as noise and misalignment of the real microscope.

Hybrid approaches have also been proposed. For example, the DeepLFM by Li \etal \cite{ref:LiXiaoxu} is based on first deconvolving the input LF image with a deconvolution method, and then on using a U-Net to super-resolve the deconvolved volume. 
This method produces volumes with good accuracy, but the computation workload and time are very high due to the use of a LF deconvolution step. 

In contrast, our proposed LFMNet explores for the first time the end-to-end training of a neural network with real confocal and aligned LF images. The network outputs reconstructed volumes at resolutions similar to those of the captured confocal data. Once integrated in a LF microscope, it yields a $720$-fold decrease in acquisition time ($72s$ for a confocal stack scan vs $100ms$ for a LF image capture) and a $30{,}000$-fold decrease in reconstruction time against conventional LF deconvolution methods. On average, it takes $1500s$ for a deconvolution-based reconstruction vs $50ms$ for a reconstruction with LFMNet ($30ms$ for the light field rectification, which can be avoided in a calibrated system, and $20ms$ to run LFMNet).

\section{Methods} \label{sec:methods}

In this section, we describe the main steps of our approach. We start with a description of our light field microscope and our design criteria in section~\ref{sec:LFMdesign}. In the Experiments section, we illustrate in detail our analysis of the chosen optical configuration and confirm the optimality of our design. In section~\ref{sec:sampleprep}, we provide details of the preparation of the samples imaged in our data set, and in sections~\ref{sec:datasetprep} and \ref{sec:alignment}, we present our procedure to acquire corresponding light field images and confocal stack scans and how to align the former to the latter. 
In section~\ref{sec:DLModel}, we describe the architecture of our neural network, its design criteria, and its training on the data set. Finally, in section~\ref{sec:PatchesVsfull} we describe the steps taken to make LFMNet a fully convolutional network.

\subsection{Designing a Light Field Microscope} \label{sec:LFMdesign}

Our LFM setup consists of a Zeiss Axio Observer microscope with a $40\times/0.9$ NA air objective. The MLA (from Flexible OKO optical) is placed in the lateral light path of the microscope and built in regular packing (orthogonal arrangement) with a focal length of $2.5mm$ and $112\mu m$ pitch. The MLA is followed by a $\text{1:1}$ relay lens (Edmund Optics Achromat Pair $100mm$ focal length) that translates the image formed by the MLA to the camera plane. Our camera is a Baumer VCXG-124M CMOS with $3.45\mu m$ square pixels. To scan the sample a motorized stage with universal mounting frame K (Zeiss) is used and a custom script written in MicroManager \cite{ref:Edelstein2010}, which allows the acquisition of our samples in an automatic manner.

\begin{figure}[t]
    \centering
    \includegraphics[width=\textwidth]{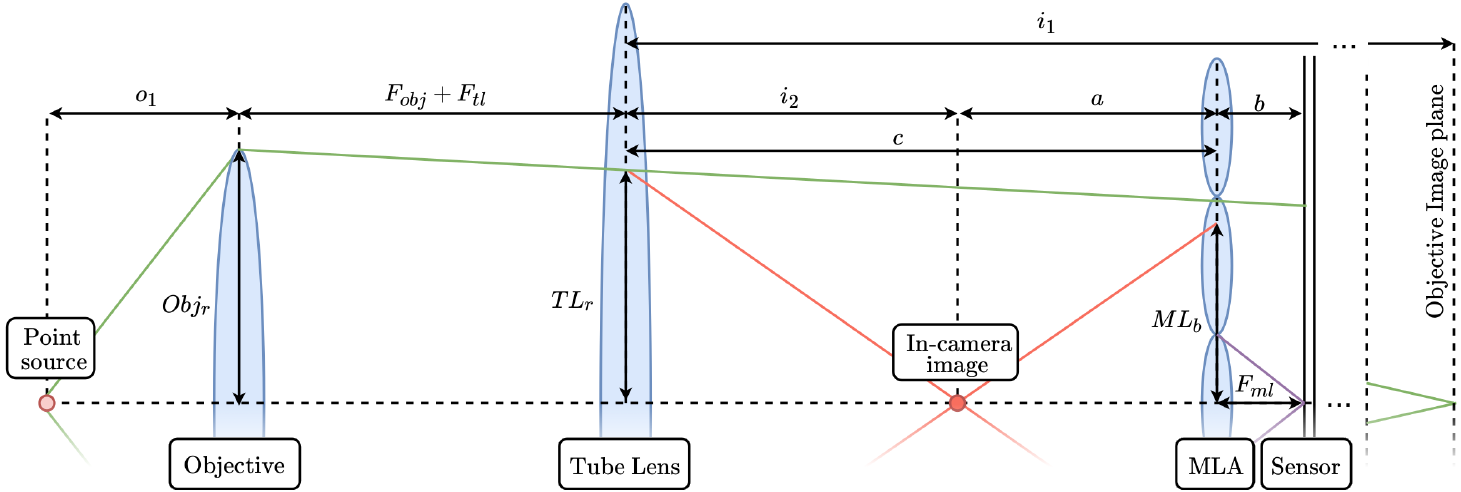}
    \caption{\textbf{Geometric optics analysis of the LFM.} This scheme illustrates the effect of the axial position of a point-light source on the blur produced at the MLA plane. A change of the object depth influences $a$ in eq.~\eqref{eq:thin_lens} because $a=\mathit{\mathit{c}} - i_2$, where $\mathit{c}$ is the distance from the tube-lens to the MLA and $i_2$ is where the intermediate image of the resolution target appears 
    (see eq.~\eqref{eq:i_2} in Appendix~\ref{sec:A:MLA_blur}).
    }
    \label{fig:MLA_Blur}
\end{figure}

The most fundamental element of the design of a LFM is the placement of the MLA relative to the imaging sensor and the object (see $a$ and $b$ in Fig.~\ref{fig:MLA_Blur}).
To identify the best settings, we evaluate a wide range of parameters against different measures of performance on simulated data. As detailed in the Experiments section, this analysis shows that \textbf{the LF-1.0 setting provides the best configuration for the depth range of interest}.
We evaluate the different configurations via a uniform grid search, where the sensor is placed at $17$ positions relative to the MLA ($b$ in eq.~\eqref{eq:thin_lens}), spanning a range from $1000\mu m$ to $5000\mu m$ with steps of $250\mu m$. We then measure the frequency response of a USAF 1951 resolution target when placed at $65$ different depths relative to the front focal plane of the objective, ranging from $-32\mu m$ to $32\mu m$ with steps of $1\mu m$. 
In this resolution target the highest frequency is at group 9, element 3 ($645.1 \text{line-pairs}/mm$ or $0.78\mu m$ line size). 
To evaluate the performance of each configuration we employ the following metrics:
\begin{enumerate}[a)]
    \item \textbf{Fisher Information (FI) of the Point Spread Function:} The FI matrix \textit{F} measures the amount of change on the sensor response, when varying the location of a source point  $\boldsymbol{p}\doteq [x_p,y_p,z_p]^\top$ 
    \begin{equation}
    \textit{F}(\textbf{p}) =
    \begin{bmatrix}
    \textit{$F_{x_p x_p}$}(\textbf{p},b) & \textit{$F_{x_p y_p}$}(\textbf{p},b) & \textit{$F_{x_p z_p}$}(\textbf{p},b) \\ 
    \textit{$F_{y_p x_p}$}(\textbf{p},b) & \textit{$F_{y_p y_p}$}(\textbf{p},b) & \textit{$F_{y_p z_p}$}(\textbf{p},b) \\  
    \textit{$F_{z_p x_p}$}(\textbf{p},b) & \textit{$F_{z_p y_p}$}(\textbf{p},b) & \textit{$F_{z_p z_p}$}(\textbf{p},b)
    \end{bmatrix}.
    \end{equation}
     A coefficient in the FI matrix is defined via
     \begin{equation}
         \textit{$F_{i j}$}(\textbf{p},b) = \iint\left ( \frac{\partial^2 \ln \widehat{h}(x,y,\boldsymbol{p},b)}{\partial i \partial j} \right ) \widehat{h}(x,y,\boldsymbol{p},b) dx dy,
     \end{equation}
     where the derivatives are taken with respect to $i,j\in\{x_p,y_p,z_p\}$, $\widehat{h}(x,y,\boldsymbol{p},b)$ is the normalized LF PSF in the image coordinates $x,y$, and with a MLA-to-sensor distance $b$.
     This performance metric was previously used by Cohen \etal \cite{ref:Cohen2014} and it measures how much information a PSF carries. One of the main advantages of this metric is that it does not depend on a specific deconvolution or reconstruction method.

    \item \textbf{USAF 1951 contrast and Pearson correlation coefficient:} This performance metrics are a measure of contrast on a resolution target. The former analyzes how the Modulation Transfer Function (MTF) varies when one places the resolution target at different depths and computes how well the line pairs can be reconstructed.
    It has been previously used as a metric for the lateral resolution \cite{ref:Broxton2013,ref:Cohen2014}. The contrast is given by ${C=(I_{max}-I_{min})/(I_{max}+I_{min})}$, where $I_{max}$ and $I_{min}$ are the maximum and minimum intensities of a patch in the observed image. The Pearson correlation coefficient \cite{Ref:corrcoeff} was computed with the Matlab Pearson correlation command \texttt{corrcoef}.
    For this experiment we place the target at the same depths as the computed PSFs from the Fisher Information experiment and compute the contrast and correlation coefficients against a ground truth resolution target (see Fig.~\ref{fig:PerfSearch}). 
\end{enumerate}

\subsection{Sample Preparation}
\label{sec:sampleprep}
\subsubsection{Fluorescent Labeling of the Vasculature}

To uniformly label all blood vessels, we inject retro-orbitally $30 \mu l$ of fluorescently labeled lectin (1mg/ml, DyLight 594conjugated, DL-1177) into isoflurane anesthetized C57BL/6J mice. After an awake period of 30 minutes, we sacrifice the mice and intracardially perfuse them with 10ml of Dulbecco's phosphate-buffered saline (PBS,14190-094) followed by perfusion with $2$\% paraformaldehyde in PBS (PFA,30525-89-4).


\subsubsection{Brain Isolation and Processing}

We carefully isolate mice brains from the skull and store them in $2$\% PFA in PBS for 16 to 20 hours post-fixation. After one day, we wash the brains with PBS and then store them for 20 to 24 hours in PBS. Then, we embed the brains in $2$\% low temperature gelling agarose (A9414) to provide tissue stability during the sectioning.
Finally, we cut $60 \mu m$ thick coronal sections of the brains using a vibrating blade microtome (Leica VT1000S) and collect the slices in PBS.


\subsubsection{Sample Mounting}

To create chambers for sample mounting that protect the brain slices from mechanical compression we glue $120 \mu m$ deep microscopy spacers (S24735) onto microscopy slides. We then position single brain slices in the center of the chamber and cover with $30 \mu l$ Mowiol mounting medium (9002-89-5). We then mount and seal the samples with a cover glass to allow for subsequent imaging. Before imaging, we store the slides in the dark at room temperature, overnight.


\subsection{Data Set Preparation}
\label{sec:datasetprep}

The next step is to build a data set containing pairs of confocal stacks and corresponding LFM images. One sample consists of a $60\mu m$ thick slice of mouse brain, where the blood vessels are fluorescently labeled with tomato lectin, with excitation and emission of $592nm$ and $617nm$ respectively. 
An important aspect of our procedure is that the LFM images and the corresponding confocal scans are aligned after capture, because the acquisitions are performed on two separate microscopes.
The following sections describe how the imaging and the alignment are performed.

\subsubsection{Confocal Microscope Image Acquisition} \label{sec:confAcqu}

We acquire the confocal microscope images on a Zeiss LSM 800 microscope. First, we image the full brain slice with a $10\times/0.45$ objective to create a coarse overview of the slice. Then, we select a region of interest (ROI) and scan it with a $40\times/1.3$ oil immersion objective. We image a grid of tiles per ROI, each spanning $1536\times1536\times64$ voxels, with a $0.087\mu m$ lateral sampling and a $0.9\mu m$ axial sampling. For the confocal data set, we acquire 3 areas, from two brain slices, of $12\times8$, $7\times17$ and $17\times12$ tiles, respectively, with a $10$\% overlap between tiles. We then mark the ROIs on the $10\times$ overview image for an easier localization of the LFM images.
To further improve the quality of the acquired volumes, we perform a deconvolution step. For this purpose, we run the Classic Maximum Likelihood Estimation algorithm with the Huygens Remote Manager, web-based implementation of Huygens Core deconvolution software (SVI, the Netherlands), for 25 iterations. Finally, the tiles are stitched together by applying a concentric gradient on the borders of each tile, and by adding them to the final image. 

\subsection{Acquisition and Alignment of LF Images and Confocal Volumes} \label{sec:alignment}

In our procedure, we use the confocal stacks as a fixed reference and adjust instead the capture and alignment of the LFM images.
To locate the LFM image within the region of interest of the volume scanned with the confocal microscope, we perform the following initial steps:
\begin{itemize}
    \item Locate the region of interest through visual inspection.
    \item Manually center the slice in $z$, such that the center of the sample is focused at the MLA.
    \item Scan a grid of $25\times25$ tiles around the detected location (the scanned grid thus covers a larger area than that of the confocal volume).
\end{itemize}
Each tile is captured by exposing the camera for $100ms$ and covers an area of $111 \times 111 \mu m$ (with $1287 \times 1287$ pixels). 
The system acquires a $25\times25$ grid of tiles covering an area of $2.77 \times 2.77 mm$ in $\sim625$ seconds (delayed mainly by the motorized stage).
\begin{figure}[t]
    \centering
    \includegraphics[width=\textwidth]{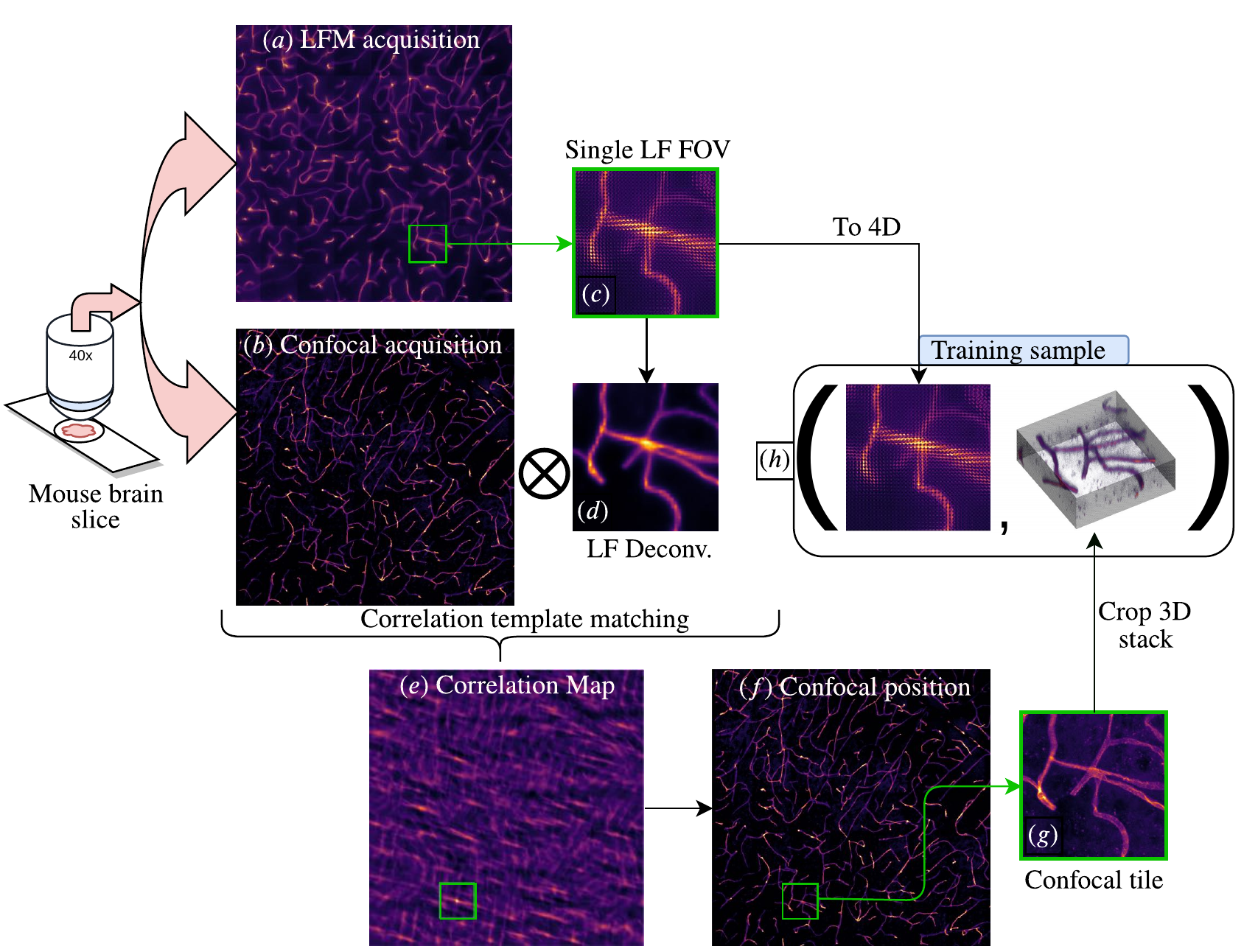}
    \caption{\textbf{Data set sample acquisition and alignment:} \textbf{(a)} LF image and \textbf{(b)} average of the confocal stack along the z axis. Both images are obtained by stitching multiple acquisitions of the same brain slice sample. \textbf{(c)} Single LF image tile to be aligned to the confocal volume. \textbf{(d)} Deconvolved \cite{ref:Stefanoiu:19} volume from the LF image in (c) averaged along the z axis. \textbf{(e)} Correlation map between (b) and (d). The region in the correlation map with the highest peak is highlighted in green. \textbf{(f)} Corresponding position of the tile found in the confocal scan. \textbf{(g)} Confocal stack crop aligned with LFM image. \textbf{(h)} The 4D LFM image and the corresponding confocal stack are then stored in the database to be used for training.}
    \label{fig:alignment}
\end{figure}
Once the LFM images (with the corresponding confocal stacks) are acquired, the following automated steps are performed to align the data (see Fig.~\ref{fig:alignment}):
\begin{itemize}
    \item \textbf{Image Rectification (Fig.~\ref{fig:alignment}~(a)).} 
    LF images are captured with a resolution of $1287\times1287$ pixels, on a grid of $39\times 39$ micro-lenses (which correspond to the spatial resolution of the LF), each covering $33\times33$ pixels (which correspond to the angular resolution of the LF). For more details on the structure of LF images see the pioneering work of Ng \etal~\cite{ref:Ng2005}.
    The LF images are rectified photometrically by using a captured white image and geometrically by using the LF Matlab Toolbox \cite{Ref:MatlabLFT}. After the geometric rectification, an integer number of pixels fits within exactly one lenslet. In our case, the original captured LFM images with $\sfrac{\text{lenslet\_pitch}}{\text{sensor\_pitch}} = 112\mu m / 3.45\mu m = 32.46$ pixels per lenslet are rectified to $33$ pixels per lenslet. This task takes $30~ms$ per image in our system.

    \item \textbf{3D Deconvolution (Fig.~\ref{fig:alignment}~(c), (d)).} Each rectified image is deconvolved into a volume that spans the same axial range as the scanned confocal stack ($64\text{ depths} \cdot 0.9\mu m = 57.6\mu m$) by using the aliasing-aware deconvolution algorithm proposed by Stefaniou \etal \cite{ref:Stefanoiu:19}. The reconstructions are sped up by using a modified implementation of the algorithm that reconstructs a smaller number of voxels (7 voxels per lenslet, which corresponds to a lateral resolution of $0.4\mu m$) per lenslet than in the original implementation. Furthermore, because the confocal microscope uses an oil immersion objective and the LFM uses instead an air objective, we compensate for the effective measured depth difference by taking the refractive index of mouse brain and immersion oil into account. We use the z-step, adjusted by the ratio $0.9/1.44$ and obtain a compensated depth range of $\sim40\mu m$.

    \item \textbf{Alignment (Fig.~\ref{fig:alignment}~(b)-(e)).} To align a LFM image tile to its corresponding confocal volume, first we average both the 3D LF deconvolution volume and confocal stack (Fig.~\ref{fig:alignment}~(b)) along their z axis. Then, we compute the correlation between these average images \cite{Ref:corrcoeff}. The method returns a 2D correlation coefficient map with a maximum peak at the highest correlated position (Fig.~\ref{fig:alignment}~(f)). To avoid false positives a peak is only selected if its value surpasses a manually chosen threshold of $0.59$ (notice that the correlation coefficient ranges between -1 and 1).

    \item \textbf{Storage (Fig.~\ref{fig:alignment}~(g), (h)).} Finally, the LFM image tile (Fig.~\ref{fig:alignment}~(h))is reshaped to a 4D LF and stored as a training sample into the data set together with the corresponding aligned confocal stack patch (Fig.~\ref{fig:alignment}~(h)).

\end{itemize}
The final data set consists of $362$ LF images composed of $33\times33\times39\times39$ elements and their corresponding confocal stacks with $1287\times1287\times64$ voxels, with voxel size $0.087\times 0.087\times 0.9 \mu m$. The data set is split so that $317$ images are used for training, $35$ for validation and $10$ for testing.


\subsection{Deep Learning Model} 
\label{sec:DLModel}


We are interested in extracting confocal stacks from LFM images. This mapping, however, presents several challenges. Firstly, since LFM images carry less data than the confocal stacks, a direct mapping is ill-posed. To introduce the missing information, we exploit the fact that the space of confocal stacks has a limited complexity and use neural networks to capture such structure. Secondly, because of the optical arrangement, LFM images do not capture the same amount of information at every depth (\ie, the effective volume slice resolution)  \cite{ref:Bishop2012a}. In particular, at the depth corresponding to a focused image on the MLA, the angular resolution is the lowest.
Using accurate models of the optics can help recover details of the reconstructed volume.
In fact, aberrations and misalignment can be beneficial, because they distort the sampling patterns defined by the ideal system and thus avoid degenerate imaging conditions.
These aspects have been exploited, for example, by Li Yi \etal \cite{ref:Wei2015}.  
However, modeling precisely aberrations, misalignment and other parameters of the optics is a very challenging task and errors can result in strong artifacts.
Thus, we address these challenges by taking advantage of the data-driven approach of deep learning (\ie, no explicit modeling is required). We train a neural network with real data and aim at learning a good prior about both the microscope setup and the data domain. 

\begin{table}[t]
    \caption{Tensor sizes for the LFMNet, whose architecture is shown in Fig.~\ref{fig:LFMNet}. $A_i$ and $S_i$ are the angular and spatial coordinates, $nD$ the number of depths and $O_i=S_i-\text{fov}+1$ the output spatial size.}
    \label{table:tens_dim}
    \centering
    \setlength{\tabcolsep}{3pt}
    \renewcommand{\arraystretch}{1}
    \begin{tabular}{|c|c||c|c|}
    \hline
    \textbf{\textit{Tensor}} & \textbf{\textit{Dimensions (ch,dim1,dim2,..)}} & \textbf{\textit{Tensor}} & \textbf{\textit{Dimensions (ch,dim1,dim2,..)}} \\
    \hline
    4D LF in & $1,A_x,A_y,S_x,S_y$ & D1 and U3& $nD,(A_x\cdot nT)/2,(A_y\cdot O_y)/2$ \\
    T1 & $nD,A_x,A_y,O_x,O_y$ & D2 and U2& $nD,(A_x\cdot O_x)/4,(A_y\cdot O_y)/4$ \\
    T2 and Vol Out & $nD,A_x\cdot O_x,A_y\cdot O_y$ & D3 and U1& $nD,(A_x\cdot O_x)/8,(A_y\cdot O_y)/8$ \\
    & & D4 & $nD,(A_x\cdot O_x)/16,(A_y\cdot O_y)/16$ \\
    \hline
    \end{tabular}
\end{table}
\begin{figure}[t]
\centering\includegraphics[width=\textwidth]{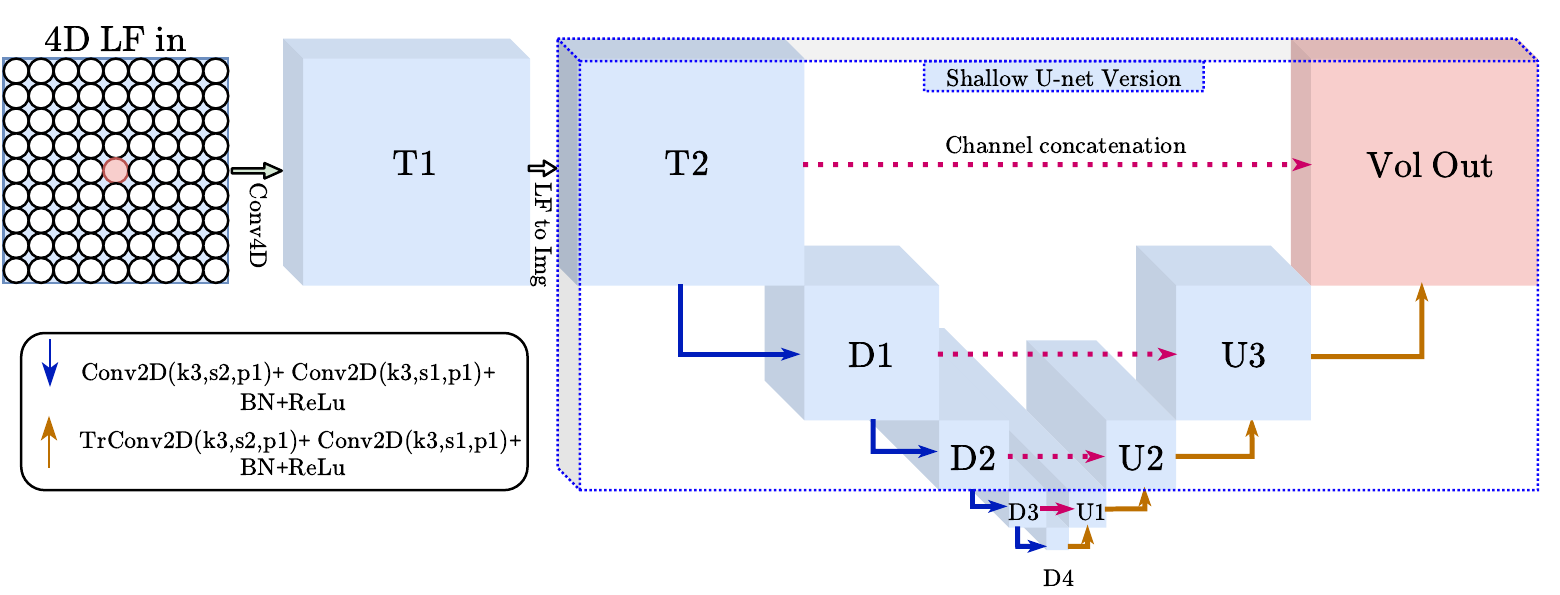}
\caption{\textbf{Proposed LFMNet architecture.} We show two different versions of the U-Net: One is the \texttt{full} version, with 4 downsampling convolutions, and the other is the \texttt{shallow} version, shown in the blue dotted rectangle. The dimensions of the tensors can be found in Table~\ref{table:tens_dim}.}
\label{fig:LFMNet}
\end{figure}

\begin{table}[t!]
    \caption{\textbf{Performance results for different LFMNet configurations.} The top part shows ablation results. The bottom part shows the final network compared with previous work. The best results per metric are in boldface. In orange we highlight the chosen network configuration. The testing is performed on 10 full LF images with shape $33 \times 33 \times 39 \times 39$.}
    \label{table:ablation}
    \centering
    \footnotesize
    \setlength{\tabcolsep}{3pt}
    \begin{tabular}{ >{\centering\arraybackslash}p{2.5cm} | >{\centering\arraybackslash}p{1.7cm} | >{\centering\arraybackslash}p{1.9cm} | >{\centering\arraybackslash}p{1.5cm} | >{\centering\arraybackslash}p{1.4cm} | >{\centering\arraybackslash}p{1.2cm} >{\centering\arraybackslash}p{1.2cm}}
    \rowcolor{lightgray!100} \textit{Model definition} & \textit{Train LF input shape} & \textit{Train volume output shape} & \textit{Full LF PSNR (db)} & \textit{Full LF SSIM} & \textit{Train time} & \textit{Full LF Test time}\\
    \rowcolor{lightgray!50} \multicolumn{5}{c}{LFMNet U-Net receptive field test (section~\ref{sec:receptiveField})} & ($sec\!\times\!\text{img}$) & ($ms\!\times\!\text{img}$)\\
    \texttt{Shallow} U-Net & 33,33,9,9 & 33,33,64 & 28.86 & 0.69 & \textbf{3.58} & \textbf{20}\\
    \texttt{Full} U-Net & 33,33,9,9 & 33,33,64 & 22.49 & 0.55 & 4.66 & 29\\
    \rowcolor{lightgray!50} \multicolumn{7}{c}{U-Net with skip connections}\\
    \texttt{Shallow} U-Net & 33,33,9,9 & 33,33,64 & 24.03 & 0.56 & 3.82 & \textbf{20}\\
    \texttt{Full} U-Net & 33,33,9,9 & 33,33,64 & 26.78 & 0.66 & 4.25 & 29\\
    \rowcolor{lightgray!50} \multicolumn{7}{c}{LFMNet Input field of view ($\text{fov}$), section~\ref{sec:FOV}}\\
    \texttt{Shallow} U-Net $\text{fov}=15$ & 33,33,15,15 & 33,33,64 & 25.61 &0.60 & 9.70 & 29\\
    \texttt{Shallow} U-Net $\text{fov}=21$  & 33,33,21,21 & 33,33,64 & 28.31 & 0.67 & 9.59 & 25\\
    \rowcolor{lightgray!50} \multicolumn{7}{c}{Number of micro-lenses to reconstruct ($nT$), with $\text{fov}$=9} \\
    \texttt{Shallow} U-Net $nT=3$ MLAs & 33,33,11,11 & 99,99,64 & \textbf{29.56} & \textbf{0.70} & 10.27 & \textbf{20}\\
    \texttt{Shallow} U-Net $nT=5$ MLAs & 33,33,13,13 & 165,165,64 & 29.12 & 0.69 & 23.97 & 23\\
    \midrule
    
    \rowcolor{lightgray!100} \multicolumn{7}{c}{Final LFMNet design and comparison to other methods (trained on 317 images)} \\
     \rowcolor{orange!50} \texttt{Shallow} U-Net, no skip conn., $\text{fov}=9$, $nT=3$ & 33,33,11,11 & 99,99,64 & \textbf{34.45} & \textbf{0.87} & 10.27 & \textbf{20}\\
         V2C Net \cite{ref:Wanga} & 33,33,39,39 & 1287,1287,64 & 28.43 & 0.76 & \textbf{0.60} & 160\\
         Deconvolution \cite{ref:Stefanoiu:19} 5~iterations & 33,33,39,39 & 1287,1287,64  & 28.64 & 0.6032 & - & $1500s$\\
    \bottomrule
    \end{tabular}
\end{table}

\subsubsection{Design Criteria} \label{sec:LFMNetdesignCriteria}
We call our proposed network LFMNet. It receives as input a LF image in the tensor format $1 \times A_x \times A_y \times S_x \times S_y $, where $S_x$ and $S_y$ are the spatial dimensions and $A_x$ and $A_y$ the angular dimensions (see also Appendix~\ref{sec:A:dimensionality}). The first layer of LFMNet is a 4D convolutional layer (more details will be presented in section~\ref{sec:PatchesVsfull}). 
The output of this layer is denoted $T1$ (see Fig.~\ref{fig:LFMNet}), with shape $nD \times A_x \times A_y \times O_x \times O_y$ where $nD$  corresponds to the number of depths of the reconstructed volume encoded in the channel dimension and $O_{x,y} = A_{x,y}-\text{fov}+1$, where the field of view \text{fov} is discussed in section~\ref{sec:PatchesVsfull}. $T1$ is then converted to a 2D image $T2$ through the mapping $nD \times A_x \times A_y \times O_x \times O_y \mapsto nD \times A_x O_x  \times A_y O_y$. Then, a modified U-Net \cite{ref:unet}, which employs convolutions as down-sampling operators, finalizes the feature extraction and 3D reconstruction. The output shape is $nD \times A_x O_x \times A_y O_y$. Furthermore, because of its fully convolutional design, our network can reconstruct the volume behind a single or multiple lenslets. The complete model is shown in Fig.~\ref{fig:LFMNet}

\subsubsection{Tiles vs Full Images and a Fully Convolutional Network} \label{sec:PatchesVsfull}

Our aim is to reconstruct the volume behind a lenslet given its neighborhood, which is a straightforward task to do when splitting the LF image into patches. To obtain a reconstruction from images with an arbitrary size, however, we need to adjust the processing of the incoming LF images so that all linear layers are convolutional. We use a 4D convolution as the input layer, which processes the spatial coordinates of the LF image micro-lens by micro-lens, and has a kernel shape just large enough to fit the neighborhood around our volume of interest. The kernel size of this layer is $3\times3\times \text{fov} \times \text{fov}$, where $\text{fov}$ is the size of the lenslet neighborhood used as input (as also discussed in section~\ref{sec:FOV}), with a padding equal to $1\times1\times0\times0$, and a stride equal to 1 in every dimension. All other layers are already convolutional and do not require an adjustment.

\subsubsection{Receptive Field Analysis} 
\label{sec:receptiveField}

\begin{figure}[t]
    \centering
    \includegraphics[width=\textwidth]{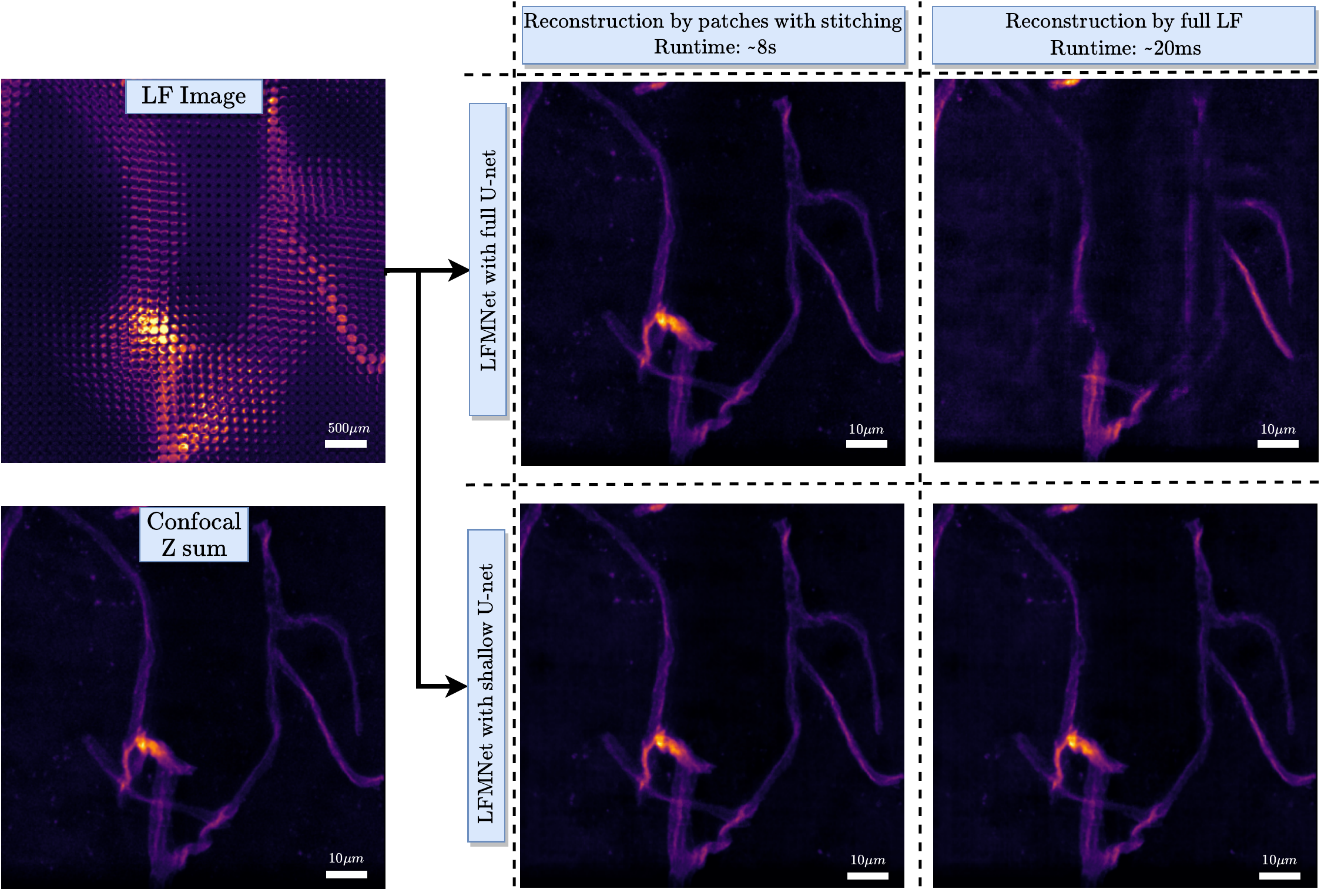}
    \caption{\textbf{Reconstruction comparison when using the \texttt{shallow} U-Net vs the \texttt{full} U-Net.} Left column: input LF image (top) and ground truth volume average $z$ projection from the confocal scan (bottom). Middle column: patch-wise reconstruction with the \texttt{full} U-Net (top) and the \texttt{shallow} U-Net (bottom). Notice that both networks produce similar results. The run-time for this reconstruction mode is $8$ seconds, which is quite slow. Right column: Fully convolutional reconstruction ($20$ ms). The reconstruction with the \texttt{full} U-Net (top) shows artifacts. Because the network is trained patch-wise, and its receptive field is larger than the patch support, it expects the input to have zero-padding beyond the patch support, and not the values found when run on the full-size input LF image. This problem is solved when limiting the receptive field to the patch support as done in the \texttt{shallow} U-Net (bottom).}
    \label{fig:u-net_comp}
\end{figure}

Another important factor for the generalization from patches to full LF images is the receptive field of the network. A single convolutional layer has a receptive field that depends on its kernel size and the stride (the down- or up-sampling factor). When stacking several convolutional layers the overall receptive field depends on their connectivity as showed by Dumoulin and Visin \cite{Ref:recField}. 
The most important aspect of this analysis is to explain how the output changes when LFMNet is fed inputs of different sizes. If the receptive field of LFMNet is larger than $A_x\times A_y$ of the input LF image region used for training, then the convolutional layers will fill in the missing data at the boundaries with reflecting conditions. However, if at test time the input LF image is larger than the receptive field of LFMNet, then the reflecting conditions are not applied and thus the network may produce an unpredictable outcome.
The narrower part of the U-Net, see Fig.~\ref{fig:u-net_comp}, contributes greatly to the broadening of the receptive field.
Thus, we consider two LFMNet implementations: one with the \texttt{full} architecture and another one without the 
las two down-up sample convolutions,
which we call \texttt{shallow} (see Fig.~\ref{fig:u-net_comp} blue dotted box).
When the receptive field covers only the central lenslet or less, as in the \texttt{shallow} architecture, the volume behind every lenslet in the LF image remains independent.
We experimented with the following settings (see Fig.~\ref{fig:LFMNet} for more details and note that the shallow LFMNet is within the blue dotted box)
\begin{enumerate}
    \item \texttt{Shallow} U-Net with 2 down/up-sample operations and a receptive field of $19$ pixels.
    \item \texttt{Full} U-Net with 4 down/up-sample operations and a receptive field of $104$ pixels.
\end{enumerate}
In Fig.~\ref{fig:u-net_comp}, we evaluate both settings under two modes of operation. In the first mode of operation, we reconstruct the full volume by processing LF patches independently (each patch has a size of $33 \times 33 \times 9 \times 9$) and then by stitching the patches together to form the output volumes. In this case, both networks produced similar results.
In the second mode, that is, when we reconstruct the full volume by processing a full LF image ($33 \times 33 \times 39 \times 39$ + padding of $4$ in the last two dimensions), only the network with the limited receptive field (the \texttt{shallow} U-Net) maintains the volume behind each lenslet untangled from its neighbors, and avoids artifacts. Moreover, this mode of operation can exploit parallel processing more efficiently than the patch-wise mode and produce the full output in just $20ms$.
Hence, we use the \texttt{shallow} U-Net for our LFMNet.

\subsubsection{Field of View of a LF Image} 
\label{sec:FOV}
The PSF of an optical system changes its support depending on the depth of the point light source. The LFMNet takes this aspect into account by using a neighborhood of lenslets to reconstruct the volume behind the central lenslets. 
Given this specification, 
we compute analytically the blur at the MLA plane generated by a point-light source placed in front of the microscope at different depths (see Fig.~\ref{fig:MLA_Blur}) according to the model used by Bishop and Favaro \cite{ref:Bishop2012a}.
The derivation of the blur-depth relationship is also reported in Appendix~\ref{sec:A:MLA_blur}.



Our analysis showed that for the desired depth range, a maximum field of view of 21 lenslets is needed (see Fig.~\ref{fig:MLBPlot} in Appendix~\ref{sec:A:MLA_blur}). In our ablation study we compare the LFMNet reconstructions when using fields of view with sizes $21\times 21$, $15\times 15$ and $9\times 9$ lenslets. We find that the $9\times 9$ fov yields the highest performance (see Table~\ref{table:ablation}). This shows that a $9\times 9$ fov is the best trade-off for our chosen architecture and training procedure between data overfitting and minimum field of view requirements. 


\begin{figure}[t]
    \centering
    \begin{subfigure}[t]{0.35\textwidth}
        \centering
        \includegraphics[height=2.4in]{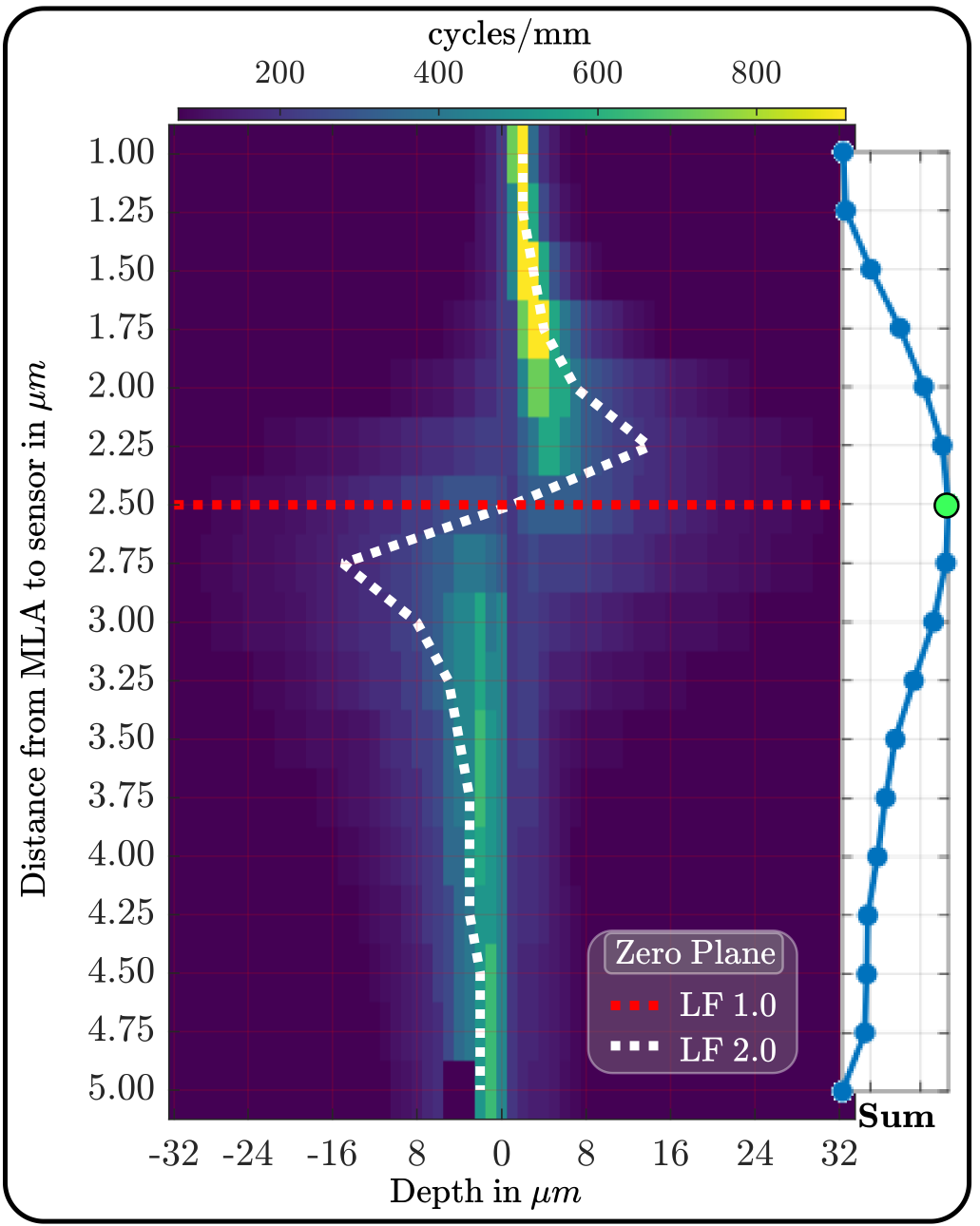}
        \caption{MTF contrast plot}
    \end{subfigure}%
    ~ 
    \begin{subfigure}[t]{0.303\textwidth}
        \centering
        \includegraphics[height=2.4in]{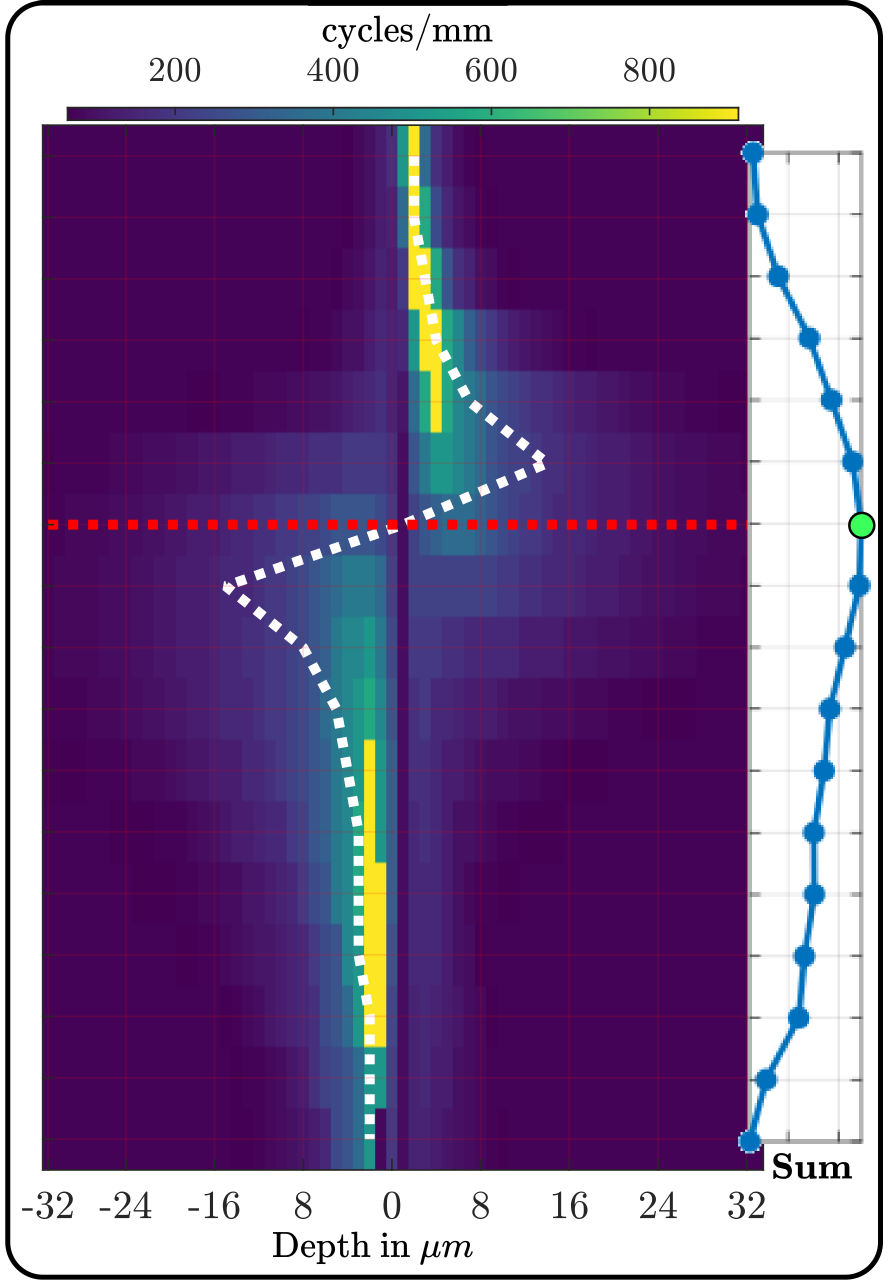}
        \caption{Correlation coefficient}
    \end{subfigure}%
    ~ 
    \begin{subfigure}[t]{0.3\textwidth}
        \centering
        \includegraphics[height=2.4in]{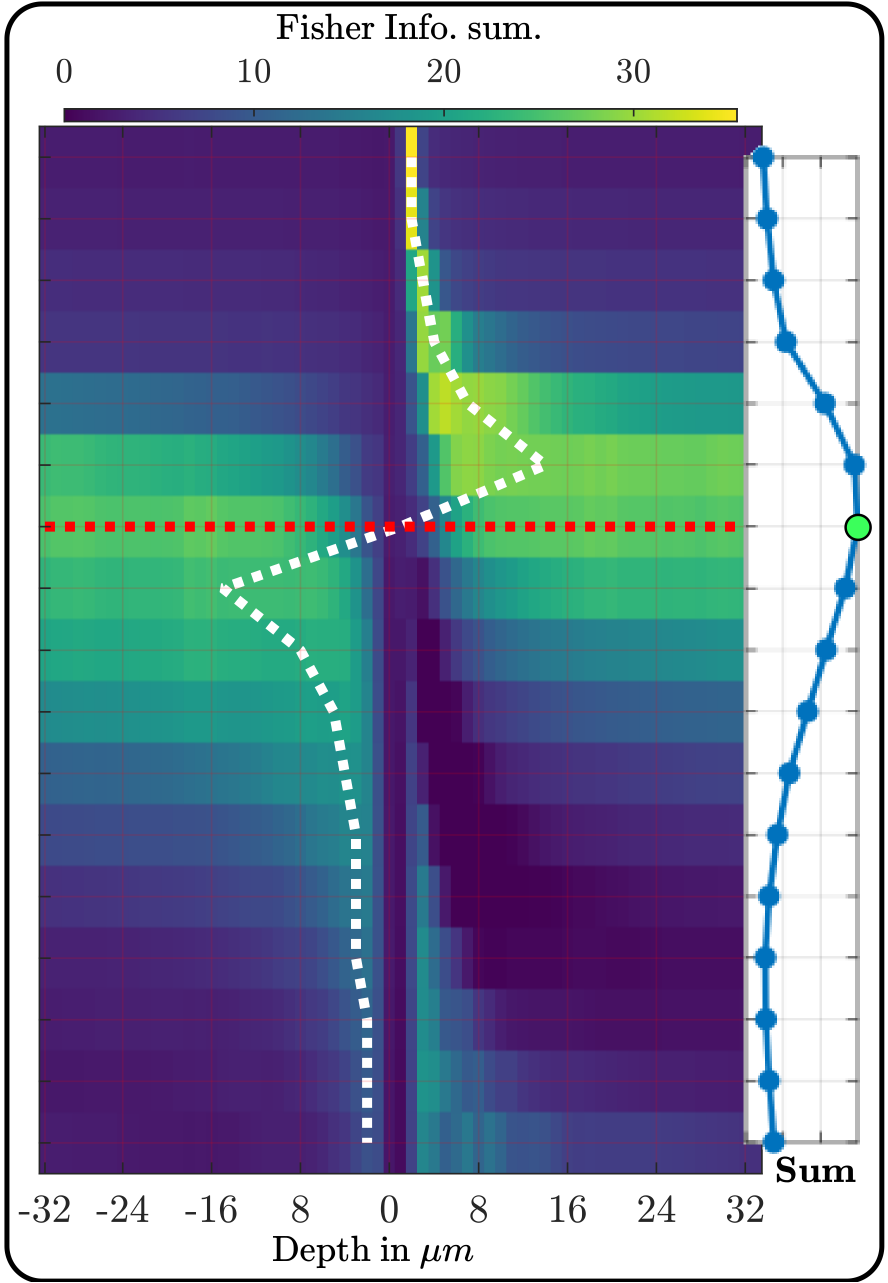}
        \caption{Fisher Information}
    \end{subfigure}%
    ~ 
    \caption{\textbf{Grid evaluation of the MLA-to-sensor distance in the LFM for different depths.} The white dotted line is the location where the focal point on a LF-2.0 LFM \cite{ref:Georgiev,ref:Lumsdaine2009} would be located (see eq.~\eqref{eq:thin_lens}). The red horizontal line shows our selected setting, equivalent to a LF-1.0 microscope. On the right side of every plot is the depth-wise sum. The green dot in each plot indicates which configuration maximizes the corresponding metric across our selected depth range.}
    \label{fig:PerfSearch}
\end{figure}

\section{Experiments} \label{sec:experiments}
\subsection{LFM Design Validation}

The analysis described in section~\ref{sec:LFMdesign} focuses on the effect of moving the sensor and the MLA within the LFM and proposes a number of performance metrics to evaluate the different configurations. 
As the first step in our analysis is to validate our model of the PSF (which is based on that of Stefanoiu \etal \cite{ref:Stefanoiu:19}) against experimental data, by looking at the Modulation Transfer Function (MTF). To do so, we first measure the MTF of our LFM setup by using the frequency responses of groups 7-9 from the USAF 1951 resolution target. Then, we compare it to the simulated MTF from our model after inserting all the calibration parameters.
The measured and simulated MTFs are shown in Fig.~\ref{fig:MTF_compare} in the Appendix~\ref{sec:A:realMTF}.
The simulated MTF (see Fig.~\ref{fig:MTF_compare}~(a)) has a slightly higher response at all frequencies than that of the measured MTF (see Fig.~\ref{fig:MTF_compare}~(b)). This mismatch is due to the inaccuracies of the LFM model of the optics, mostly caused by the relay optics. 
The important factor in this analysis is the matching focal plane between the simulated and the real MTF.

After validating our PSF model, we can evaluate the LFM setup settings. In Fig.~\ref{fig:PerfSearch}, we show each performance metric for a wide range of settings in the form of a heat-map: For each setting the heat-color corresponds to the highest frequency with at least $80$\% of (a) contrast, (b) correlation and (c) Fisher Information. Thus, bright (yellow) colors correspond to high frequencies, which should yield high resolution reconstructions, and dark (blue) colors correspond to low frequencies, which imply a poorer reconstruction quality.
We also highlight with a white dotted line the optimal MLA distance for each object depth according to the LF-2.0 design \cite{ref:Georgiev,ref:Lumsdaine2009}.
Note how the LF-2.0 setting follows approximately the profile with the highest peak (\ie, best high-resolution reconstruction) for each MLA distance.
This is particularly useful when imaging thin volumes, but this is not our case, as our setup acquires a volume spanning $57.6\mu m$.
On the right-hand side of each heat-map in Fig.~\ref{fig:PerfSearch}~(a)-(c), we plot the average performance across the depths for each MLA placement. Since all the chosen metrics yield a high value when the performance is high, the maximum value across all object depths (marked with a green dot) indicates the best MLA setting for our LFM. This choice matches the conventional LF-1.0 design \cite{ref:Georgiev,ref:Lumsdaine2009}, and thus we used it for our final set up.

\subsection{Network Ablation and Performance Evaluation} \label{sec:ablation}

\begin{figure}
    \centering
    \includegraphics[width=0.95\textwidth]{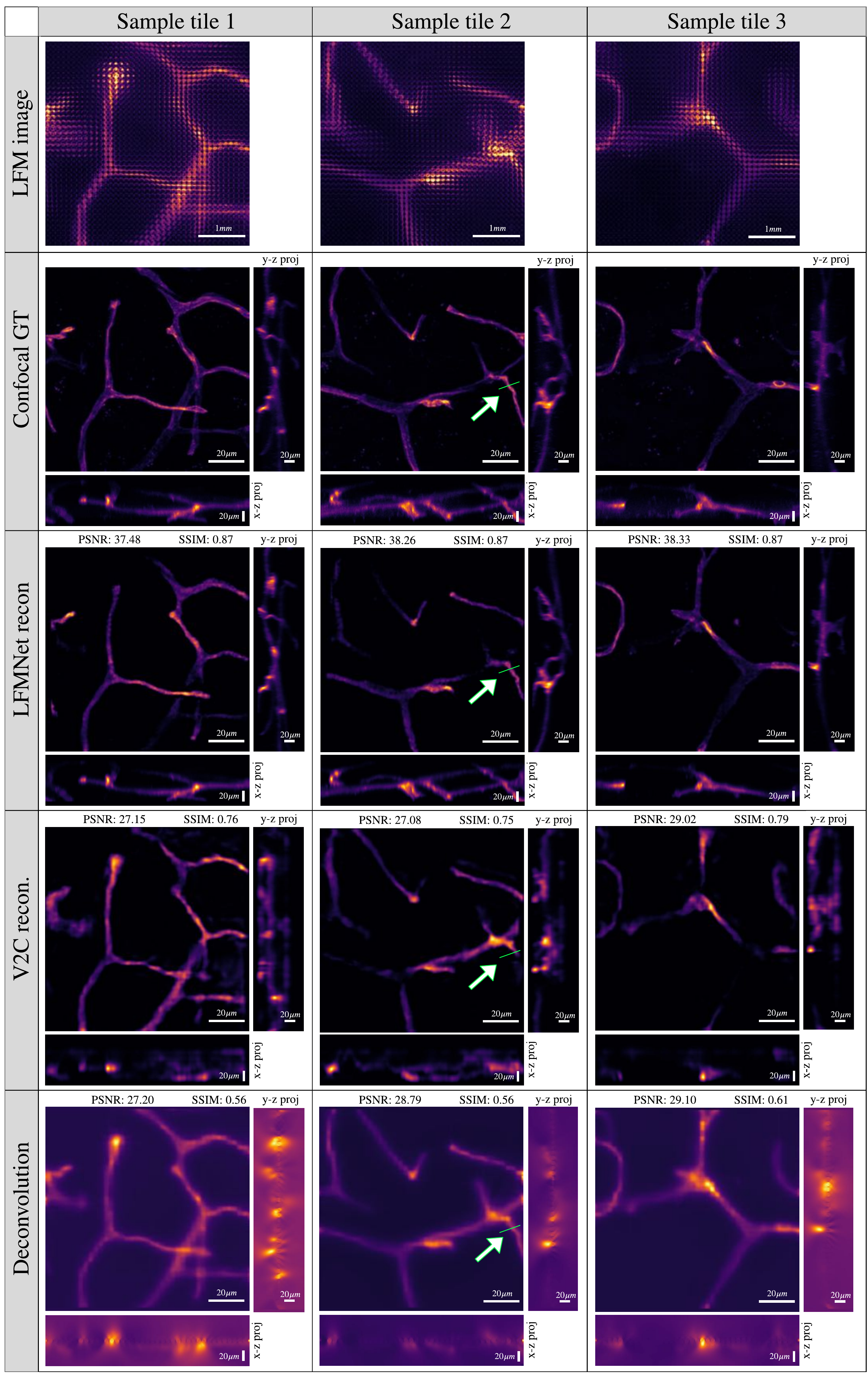}
    \caption{\textbf{Reconstruction comparison between the proposed LFMNet, the V2C network \cite{ref:Wanga} and LF deconvolution \cite{ref:Stefanoiu:19}.} Each column corresponds to a different sample. The top row is the input LF image. The second row is the ground truth from the confocal microscope. The third row is our reconstruction, followed by the V2C and the deconvolution reconstructions. Direct visual inspection reveals a drastic difference in accuracy between our reconstruction and that of prior work. For visualization purposes, all images show only the top $95$\% of the intensity, so that the background auto-fluorescence is not visible. Note the green line in sample 2, for which a profile and further analysis is shown in Fig.~\ref{fig:veinDetail}.}
    \label{fig:ReconComparison}
\end{figure}

In this section we evaluate the effectiveness of our method on the design criteria introduced in section~\ref{sec:DLModel}, by comparing side-by-side the two U-Net configurations (\texttt{shallow} and \texttt{full}), the use of skip connections in the U-Net architecture, the different field of view choices (fov) and the volume size used for training, denoted by $A_x nT \times A_y nT \times nD$, where $nT \in \{3,5\}$.
The networks in Table~\ref{table:ablation} are trained on a subset of the data formed by $27$ images ($41{,}067$ patches), and validated on $3$ images ($4{,}563$ patches). Then, we use a subset of $10$ full LF images ($33\times 33\times 39\times 39$) for testing. The quality measures of a reconstruction are the Peak Signal to Noise Ratio ($PSNR$) and the Structural Similarity Index ($SSIM$) \cite{ref:ssim}. Also, we show the training and reconstruction times for a single LF image.
As a result of our ablation, the design that yielded the highest image PSNR and SSIM (see 4th and 5th columns in Table~\ref{table:ablation}) is
\begin{quote}
\emph{
    The \texttt{shallow} U-Net without skip connections, trained with $33\times33\times11\times11$ input LF patches and with an output volume of~  $99\times99\times64$ voxels. This corresponds to the volume behind $3\times 3$ micro-lenses ($nT$), given a $\text{fov}=9$ per micro-lens.}
\end{quote}
Next, we compare the LFMNet with this settings (highlighted in orange on the bottom of Table~\ref{table:ablation}) against the V2C approach \cite{ref:Wanga} and the aliasing-aware deconvolution \cite{ref:Stefanoiu:19}. Both networks are trained with $370$ LF images (LFMNet with real LF images and V2C net with simulated LF images), and tested on the $10$ LF images from the separate test set. 
LF images reconstructed with the LFMNet, V2C and deconvolution are qualitatively compared in Fig.~\ref{fig:ReconComparison}. A slice through a single vessel is depicted in Fig.~\ref{fig:veinDetail}.  Our proposed method provides overall a $75{,}000$ fold improvement in reconstruction time against deconvolution, and a higher reconstruction accuracy in terms of PSNR and SSIM.

\subsection{Resolution Analysis} \label{sec:resolution}
\begin{figure}[t]
    \centering
    \begin{subfigure}[t]{0.24\textwidth}
        \centering
        \includegraphics[width=\linewidth]{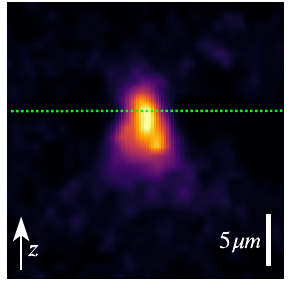}
        \caption{Confocal}
    \end{subfigure}
    \begin{subfigure}[t]{0.24\textwidth}
        \centering
        \includegraphics[width=\linewidth]{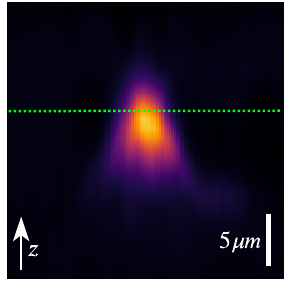}
        \caption{LFMNet}
    \end{subfigure}
    \begin{subfigure}[t]{0.24\textwidth}
        \centering
        \includegraphics[width=\linewidth]{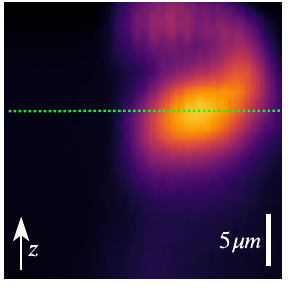}
        \caption{V2C Net \cite{ref:Wanga}}
    \end{subfigure}
    \begin{subfigure}[t]{0.24\textwidth}
        \centering
        \includegraphics[width=\linewidth]{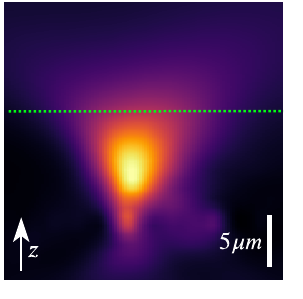}
        \caption{LF Deconv. \cite{ref:Stefanoiu:19}}
    \end{subfigure}\\
    \hspace{-3cm}
    \begin{subfigure}[b]{0.6\textwidth}
        \centering
        \includegraphics[width=\linewidth]{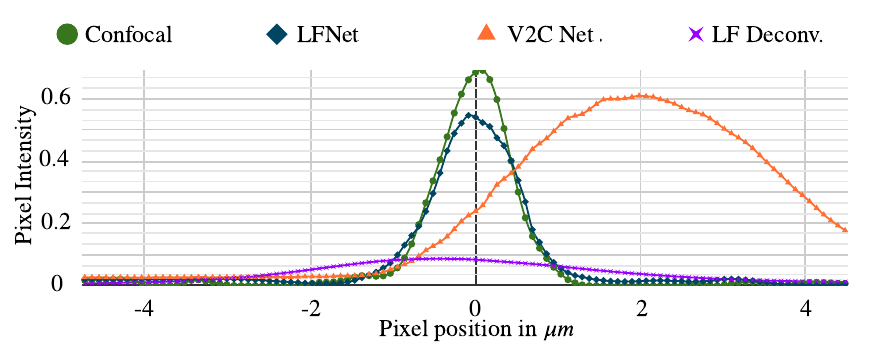}
        \caption{Intensity profile from the green line in (a)-(d)}
    \end{subfigure}
    \begin{subfigure}[b]{0.2\textwidth}
        \centering
        \begin{tabular}{c|c}
            \textbf{Method} & \textbf{FWHM ($\mu m$)} \\
            \hline
            Confocal & 1.032 \\
            LFMNet & 1.118 \\
            V2C\cite{ref:Wanga} & 3.698 \\
            LF Deconv. \cite{ref:Broxton2013} & 3.950
        \end{tabular}
        \caption{}
    \end{subfigure}

    \caption{\textbf{Brain vessel axial image comparison}. \textbf{(a)-(d)} show the projection of a blood vessel  with different reconstruction methods. The projections are taken from the green line (and indicated by the white arrow) shown in Fig.~\ref{fig:ReconComparison}, Sample~tile~2. \textbf{(e)} shows the intensity profiles through the middle of the blood vessel (green dotted line in (a)-(d)) of different methods. \textbf{(f)} shows the full width at half maximum comparison.}
    \label{fig:veinDetail}
\end{figure}

Measuring the spatial resolution of a microscope is commonly performed by imaging fluorescent beads smaller than the resolution limit \cite{ref:Wanga, ref:Broxton2013}. This strategy is not possible in our method, as the LFMNet learns a strong data prior, which makes it incapable of reconstructing samples not present in the training set (\eg, micro-spheres and resolution targets). Instead, we measure the 3D profile of a blood vessel and compare the reconstructions with different methods. In Fig.~\ref{fig:veinDetail} we show a projection of a 3D slice from a blood vessel indicated by the white arrow in Fig.~\ref{fig:ReconComparison}, Sample~tile~2. Also, the table in Fig.~\ref{fig:veinDetail}~(f) compares the Full Width at Half Maximum (FWHM) of an intensity profile across these projections. We show that LFMNet can resolve a blood vessel with $0.086 \mu m$ error, in contrast to the V2C and LF deconvolution methods, which obtain $2.666\mu m$ and $2.918\mu m$ errors.


\subsection{Data Set Evaluation}

To better understand the nature and structure of the created data set, we provide some measure of the variability and complexity of the captured samples. 
Rather than computing statistics of the raw data (\ie, the pixel intensities), we use the first two layers of a pre-trained VGG-16 \cite{ref:JohnsonAL16} as feature extractors.
Then, we compare the statistics of these features to those of other data sets. We consider three reference data sets: One is ImageNet \cite{ref:imagenet}, which contains images with high texture diversity. A second data set is the C-elegans, which consists of confocal stacks \cite{ref:epfl}, and a third data set is CCDB, which consists of mouse neuronal data \cite{ref:CCDB}.

\begin{figure}[t]
    \centering
    \begin{subfigure}[t]{0.45\textwidth}
        \centering
        \includegraphics[width=\linewidth]{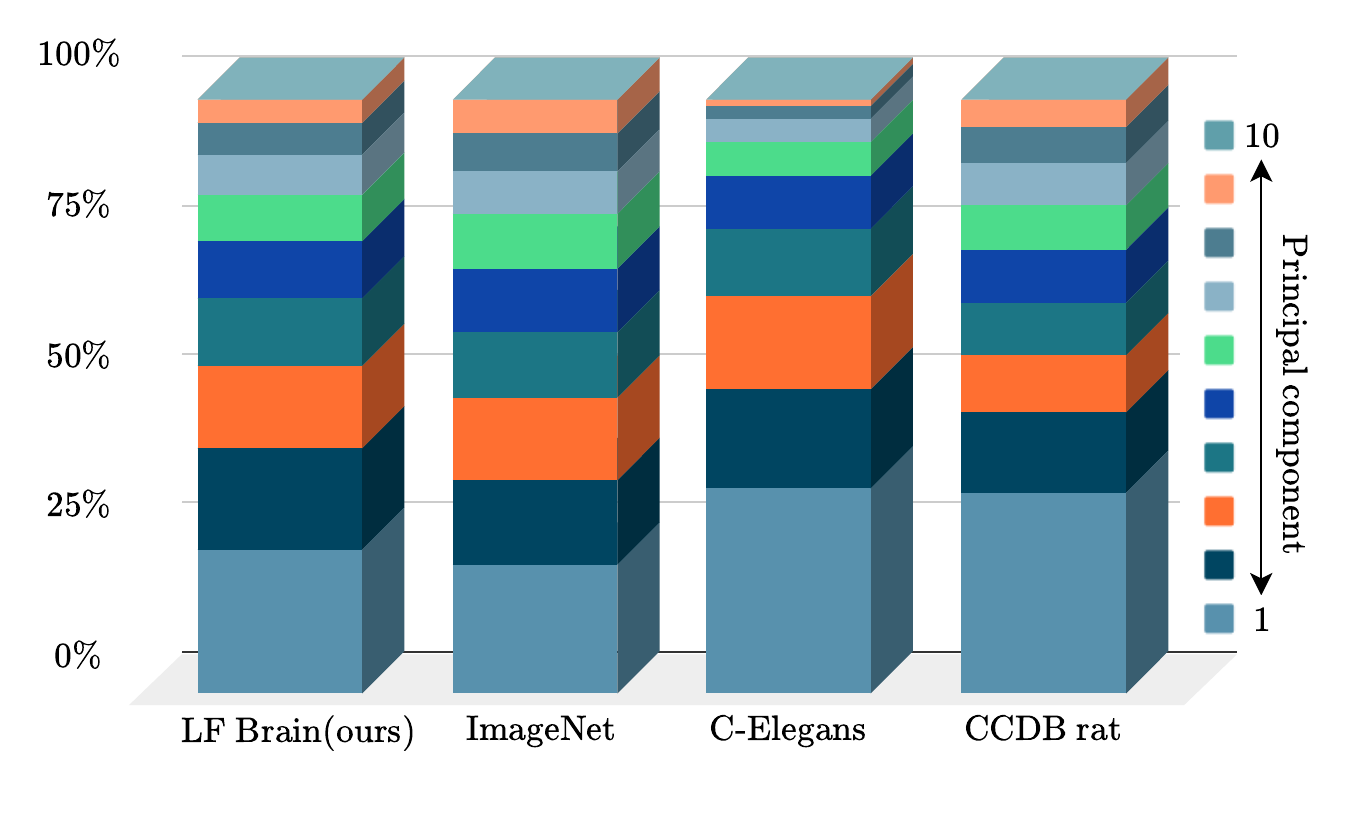}
        \caption{}
    \end{subfigure}
    \begin{subfigure}[t]{0.45\textwidth}
        \centering
        \includegraphics[width=\linewidth]{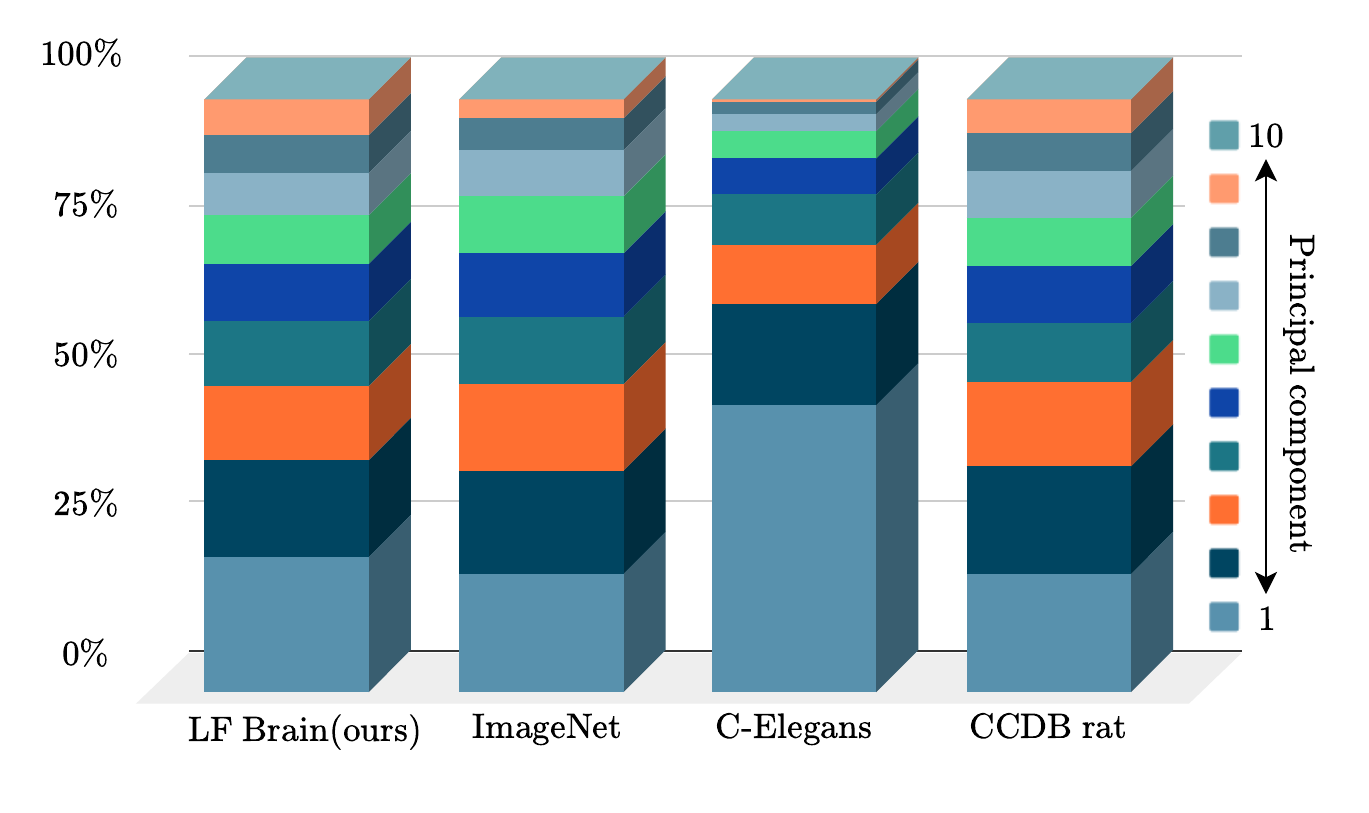}
        \caption{}
    \end{subfigure}\\\vspace{0.5cm}
    \begin{subfigure}[t]{0.45\textwidth}
        \centering
        \includegraphics[width=\linewidth]{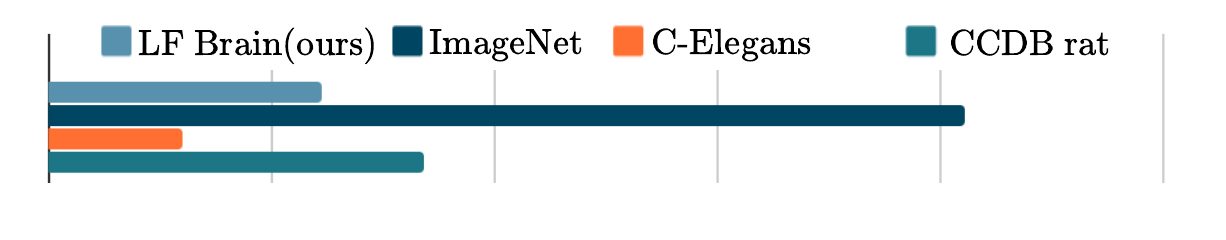}
        \caption{}
    \end{subfigure}
    \begin{subfigure}[t]{0.45\textwidth}
        \centering
        \includegraphics[width=\linewidth]{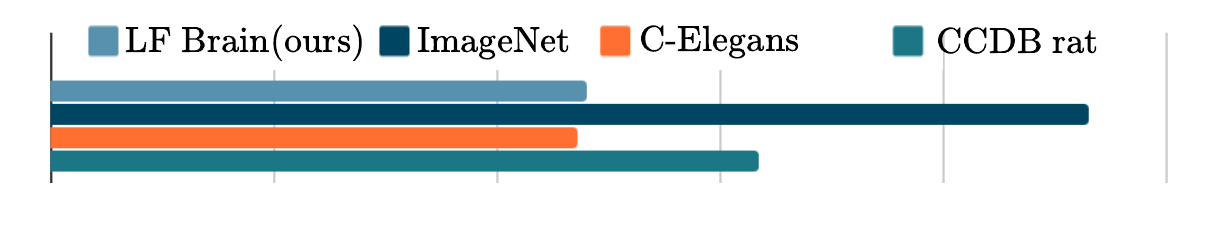}
        \caption{}
    \end{subfigure}
    \caption{\textbf{Statistical comparison of data sets.} (a) and (b): Block histograms showing the relative importance of each PCA coefficient, on VGG-16 relu1 (left) and relu2 (right) features. (c) and (d): Coefficients of variation of VGG-16 relu 1 (left) and relu 2 (right).}
    \label{fig:Datasets}
\end{figure}

We use the first 10 components of the Principal Component Analysis (PCA) and the coefficient of variation (\sfrac{$\sigma$}{$\hat{x}$}) of the features as measures of the data complexity.
As one can see in Fig.~\ref{fig:Datasets}, the complexity in terms of both the PCA and the coefficient of variation of our LF brain data set sits between that of ImageNet and that of both CCDB and C-elegans. 
As expected, ImageNet has a large coefficient of variation, but our data set is quite similar in complexity to other useful data sets in biology. 

\section{Discussion}
\label{sec:discuss}

The optical components used in our LFM perform a high density sampling of the spatial and angular information of the real world light field, due to the $112\mu m$ lenslets diameter and the $3.45\mu m$ camera pixel width. These reduced dimensions of the pixels were chosen to enable the reconstruction of the object at a resolution close to that of a confocal scan when using a deep learning approach ($0.086\mu m$ error when reconstructing a blood vessel).
However, small pixels also have a poor signal to noise ratio. To compensate for the noise, these sensors require a high exposure time and thus have a limitation in the achievable frame rate when used to capture videos. This trade-off between the temporal and the angular resolution should be taken into consideration when designing LF microscopes. In our case, we find that 10 frames per second is a frame rate that makes the scanning of large volumes, such as a whole mouse brain, practical (about 30 minutes). 
We leave the design of a LFM capable of real-time \textit{in vivo} imaging to future work. For example, one could employ a modern sCMOS camera with high quantum efficiency, high frame-rate and larger pixel size (\eg {$6.5 \mu m $ or $4.25 \mu m$}) to achieve shorter exposure times, while maintaining a high signal to noise ratio. 

More in general, the design of the whole system would benefit from a computational photography approach, where the microscope setup and the reconstruction network are jointly optimized. This could yield a novel design of the optics that would allow the microscope to capture more information per pixel, by sacrificing the interpretability of the captured image.
This form of \emph{optical image compression}, as already demonstrated by our LFM, would enable faster scan times and a smaller data storage than existing methods (such as light-sheet, confocal or multi-photon microscopy). The reconstruction network would work as a decompression algorithm that recovers the full volume scan. We expect that such a system would achieve an even higher performance than what we have demonstrated with our LFMNet.

\section{Conclusion} 

We introduce an alternative to confocal microscopy that is faster and more storage efficient, while not compromising the accuracy of the volume scan. Our proposed solution consists of a light field microscope (LFM) combined with LFMNet, a novel neural network to reconstruct volumes from light field images. We showed how we chose the LFM settings by using contrast, correlation coefficient and Fisher Information (FI) as performance metrics on our model of the microscope point spread function (PSF), and have validated our analytical model of the PSF experimentally. 
We found all these metrics provide similar insights, as also previously noticed by Cohen \etal \cite{ref:Cohen2014}. Thus, we recommend the use of the lone FI as a metric as it is the most computationally efficient.
We also introduced an analysis of the LFMNet architecture and showed how the network can train on light field patches, but then be used on larger light field images in the testing phase.
As shown on our test data, LFMNet yields state of the art results and an accurate volume reconstructions with confocal-level details. To train LFMNet we also built a data set of mouse brain blood vessels with light field image and volume scan pair samples. The data set and the source code will be publicly available for research purposes \cite{ref:MiceLFMDataset}.


\section*{Funding}

 This project was financially supported by the interdisciplinary project funding UniBE ID Grant 2018 of the University of Bern.

\section*{Disclosures}

The authors declare no conflicts of interest.

\appendix
\appendixpage

\section{Format and Representation of LF Data} \label{sec:A:dimensionality}

A light field is a 4D function with dimensions $S_x \times S_y \times A_x \times A_y$, where the first two coordinates sample the spatial domain and the last two coordinates sample the angular domain. The 2D spatial coordinates define a position in the object space. The 2D angular coordinates define instead the angle of the ray (in the geometric optics approximation) from which the object is observed. 
The 2D camera sensor of the LFM captures an image of $A_x S_x \times A_y S_y$ pixels, which we call the \emph{spatial representation} of a LF.
To map the LF, which is a 4D function, to the sensor, which is 2D, a light field microscope uses a micro-lens array inserted as shown in Fig.~\ref{fig:MLA_Blur}. 
The micro-lens array has $S_x \times S_y$ micro-lenses and is aligned to the sensor so that each micro-lens covers a region of $A_x \times A_y$ pixels.  
In our setup we set $S_x=S_y=39$ micro-lenses (imaged by the whole sensor) and $A_x=A_y=33$ pixels (per micro-lens). Notice that $A_x$ and $A_y$ can be obtained approximately by dividing the micro-lens diameter by the pixel width, \ie, $112\mu m / 3.45\mu m \simeq 33$~pixels.
\begin{figure}[t]
    \centering
    \begin{subfigure}[t]{0.33\textwidth}
        \centering
        \includegraphics[height=1.6in]{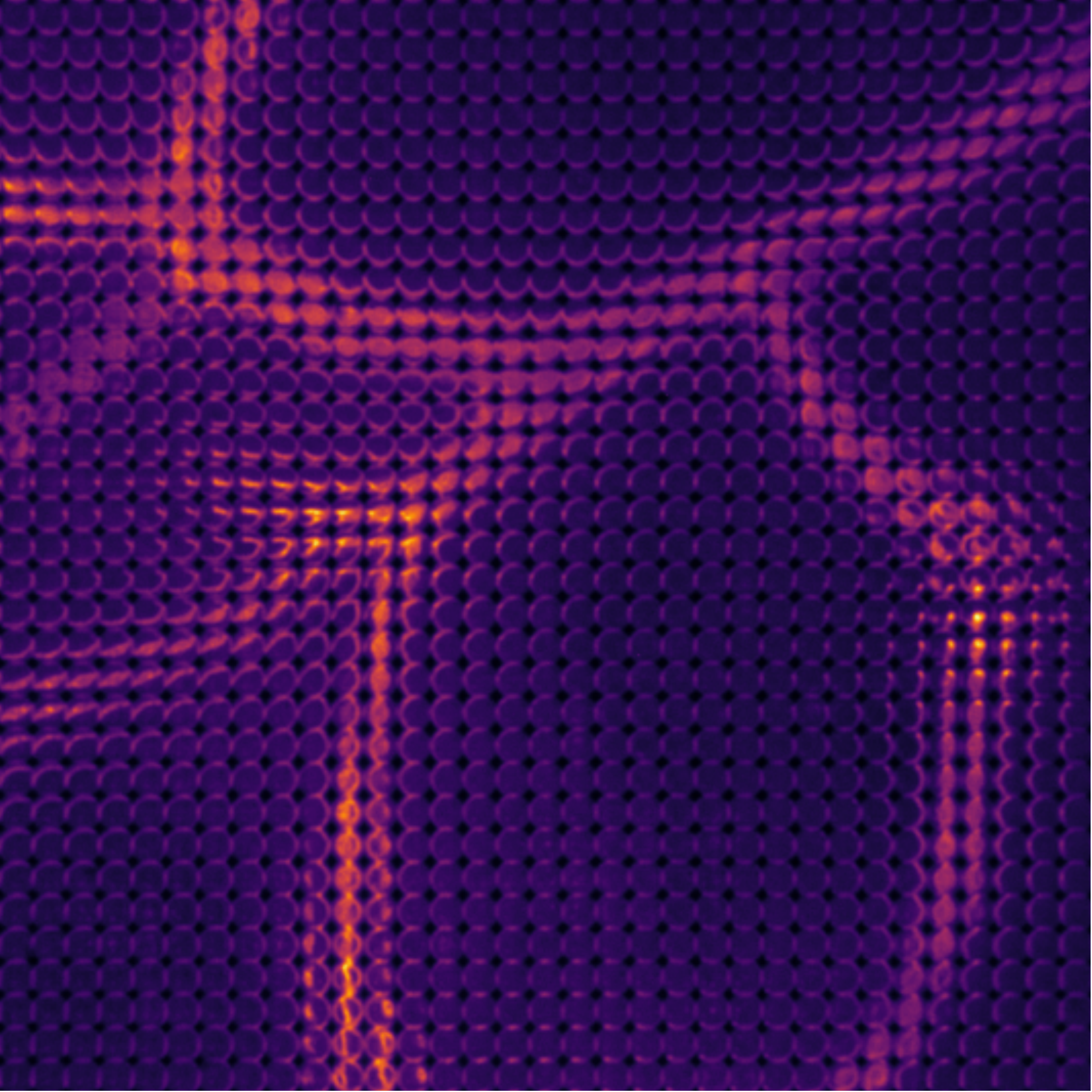}
        \caption{}
    \end{subfigure}%
    ~ 
    \begin{subfigure}[t]{0.33\textwidth}
        \centering
        \includegraphics[height=1.6in]{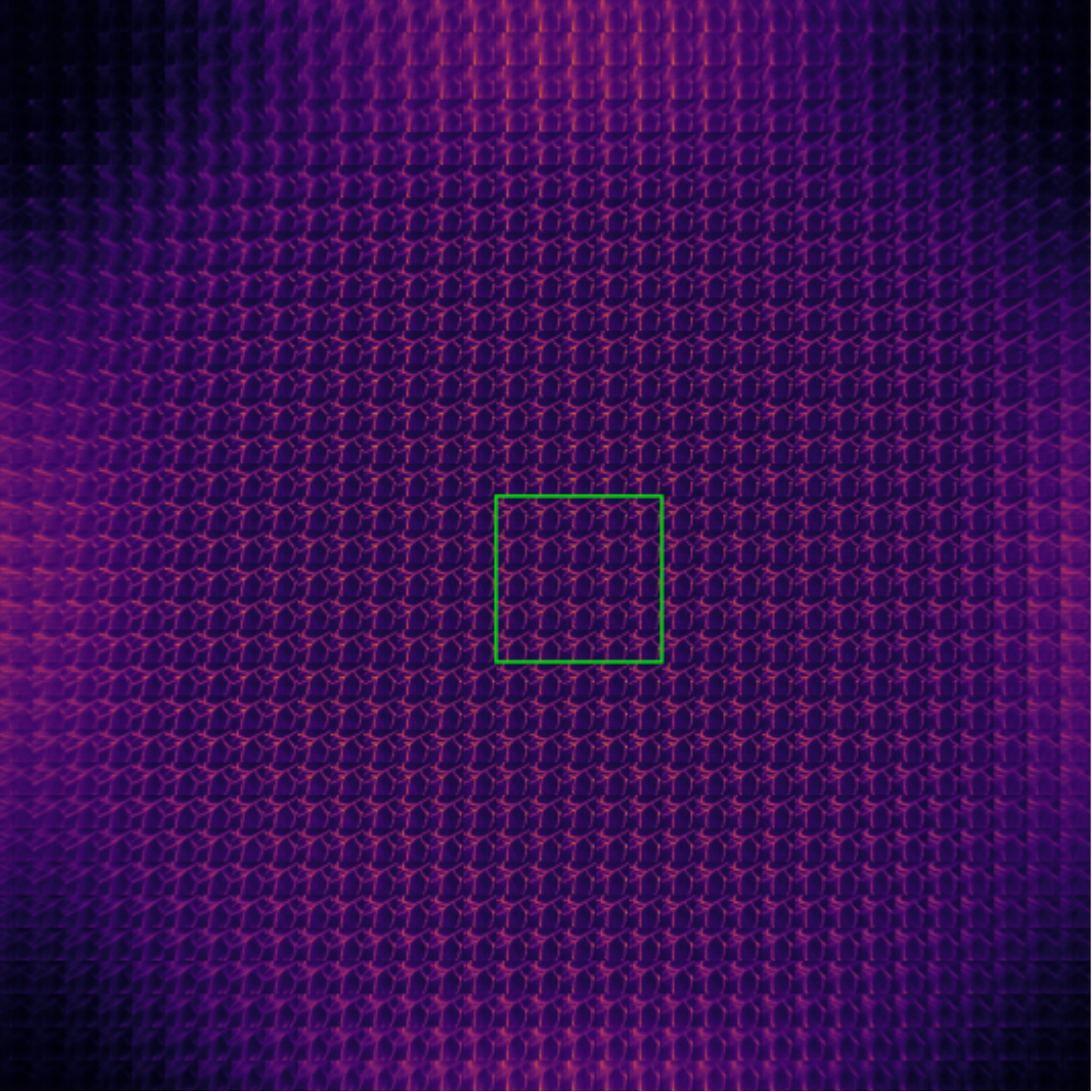}
        \caption{}
    \end{subfigure}
    ~
    \begin{subfigure}[t]{0.29\textwidth}
        \centering
        \includegraphics[height=1.6in,cfbox=green 1pt 1pt]{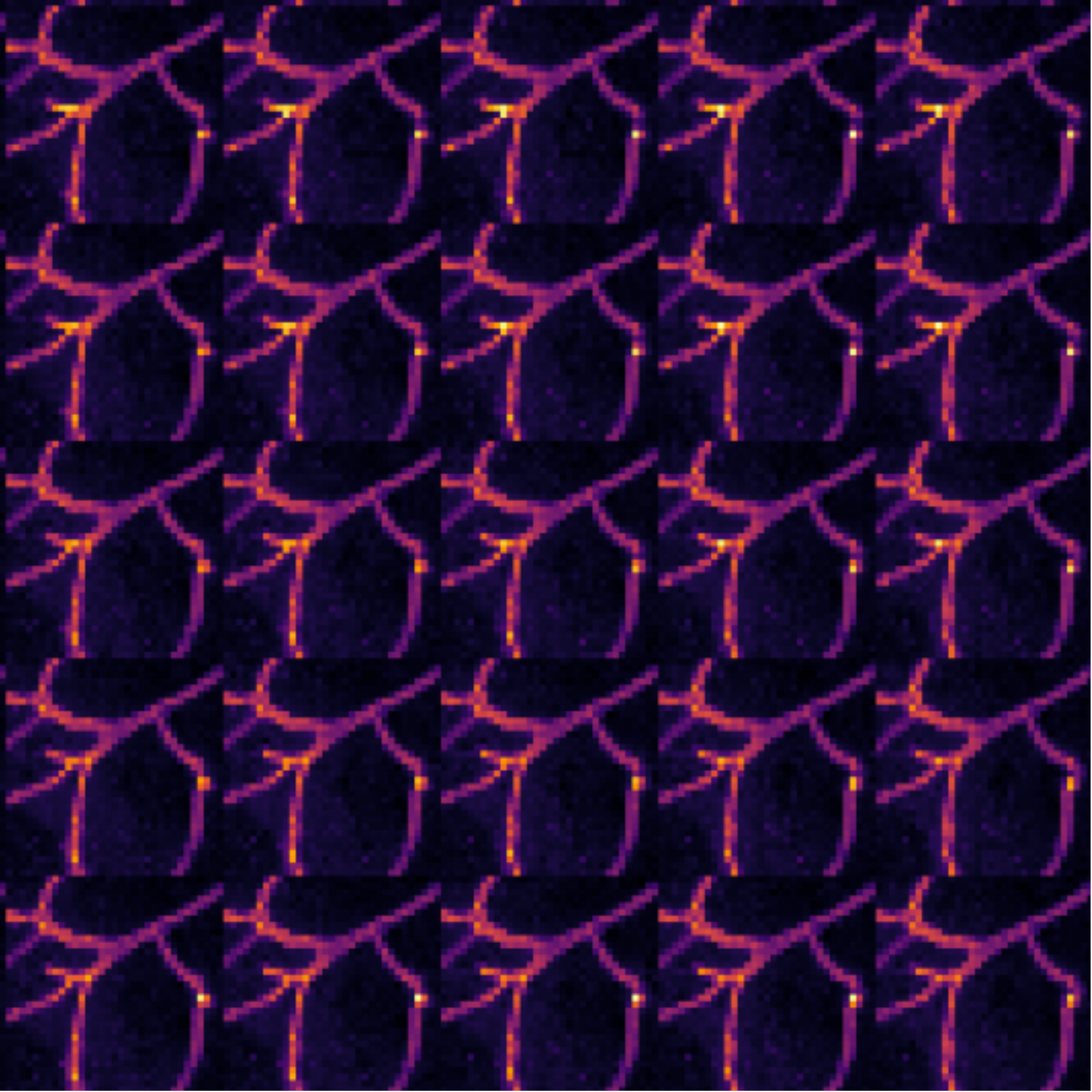}
        \caption{}
    \end{subfigure}
    \caption{\textbf{Spatial and angular LF representations.} (a) Spatial representation of the LF (raw LFM image). (b) Angular representation of the LF. (c) Magnification of the region shown in (b). The number of micro-lenses is the number of pixels in each subimage.}
    \label{fig:LFDims}
\end{figure}
The LF 4D information can be embedded and visualized as a 2D image in two ways
\begin{itemize}
    \item \textbf{Spatial representation.} This is the format with which the LF is acquired by the LFM, where each micro-lens samples the object from a different spatial coordinate, as shown in Fig.~\ref{fig:LFDims}~(a), and every pixel inside a micro-lens captures light from the object along different angles.
    \item \textbf{Angular representation or perspective views.} This arrangement is achieved by gathering all the pixels that are located at the same distance relative to each micro-lens center. Because each pixel behind a micro-lens gathers information coming from a different angle, the angular representation is somehow analogous to a camera array view (which we also call \emph{perspective view}) where each camera captures a small image from a different angle. These views are tiled as shown in Fig.~\ref{fig:LFDims}~(b) and in the enlargement Fig.~\ref{fig:LFDims}~(c).
\end{itemize}

\begin{figure}
    \centering
    \includegraphics[width=0.5\textwidth]{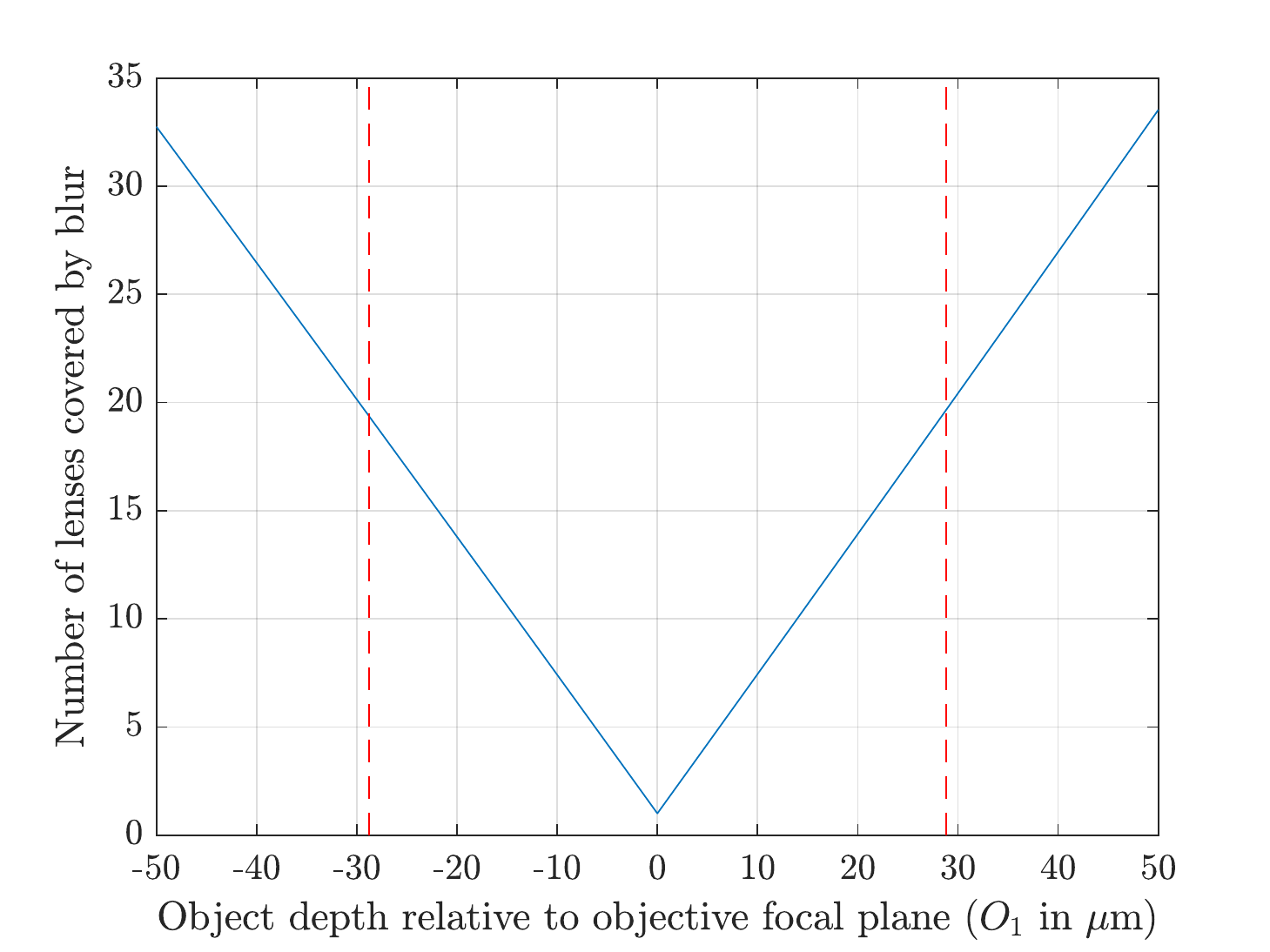}
    \caption{\textbf{Blur at MLA against number of lenslets.} Number of micro-lenses overlapping with the blur from a object with size $O_s= \frac{MLpitch}{M} = \frac{112}{40}\mu m = 2.8 \mu m$ placed at different depths ($O_1$) in front of the microscope. The red lines show the range of depths used in our setup (-28.8 to 28.8 $\mu m$). In this range the blur covers approximately 20 micro-lenses. }
    \label{fig:MLBPlot}
\end{figure}

\section{Blur of an Object at the MLA Plane} \label{sec:A:MLA_blur}
As described in section~\ref{sec:FOV} the number of MLAs that gather information of an object in front of the microscope depend on the MLA blur equation, given by
\begin{equation}
    ML_b = \frac{TL_r \cdot|\mathit{c}-i_2|}{i_2},
\end{equation}
where $TL_r$ is the blur radius size at the tube-lens, $\mathit{c}$ the distance from the tube-lens to the MLA and $i_2$ the position where the image is formed by a point in object space. We take into account that the objective back aperture (with radius $Obj_r=F_{obj} \cdot NA$) works as a telecentric stop, and that the distance between the objective and the tube-lens is equal to the sum of their focal lengths (as in a 4-F system), which is equal to $(M+1)\cdot F_{obj}$. 
From similar triangles, we find that
\begin{equation}
    TL_r = \frac{Obj_r[i_1-(M+1)F_{obj}]}{i_1}.
\end{equation}
When imaging an object of size $Os$, its blur diameter is
\begin{equation}
    ML_{tb} = 2\cdot ML_b + M\cdot Os.
    \label{eq:ml_tb}
\end{equation}
By using the thin lens equation $\sfrac{1}{i}+\sfrac{1}{o} = \sfrac{1}{F}$, we can express $i_1$ and $i_2$ in terms of the point emitter and the position in object space $o_1$ such that
\begin{equation}
    i_1 = \frac{F_{obj} o_1}{o_1-F_{obj}}\\
    i_2 = M\big(F_{obj}(1+M) - M\cdot o_1\big).
    \label{eq:i_2}
\end{equation}
This relationship between object depth and MLA blur can be better observed in Fig.~\ref{fig:MLBPlot}.
Even though the extent of the PSF at the MLA with our setup theoretically covers $\sim20$ micro-lenses, having such a large input (and 4D convolution kernel) hampers the training time considerably and might incur more easily overfitting.


\begin{figure}[t]
    \centering
    \begin{subfigure}[t]{\textwidth}
        \centering
        \includegraphics[width=\textwidth]{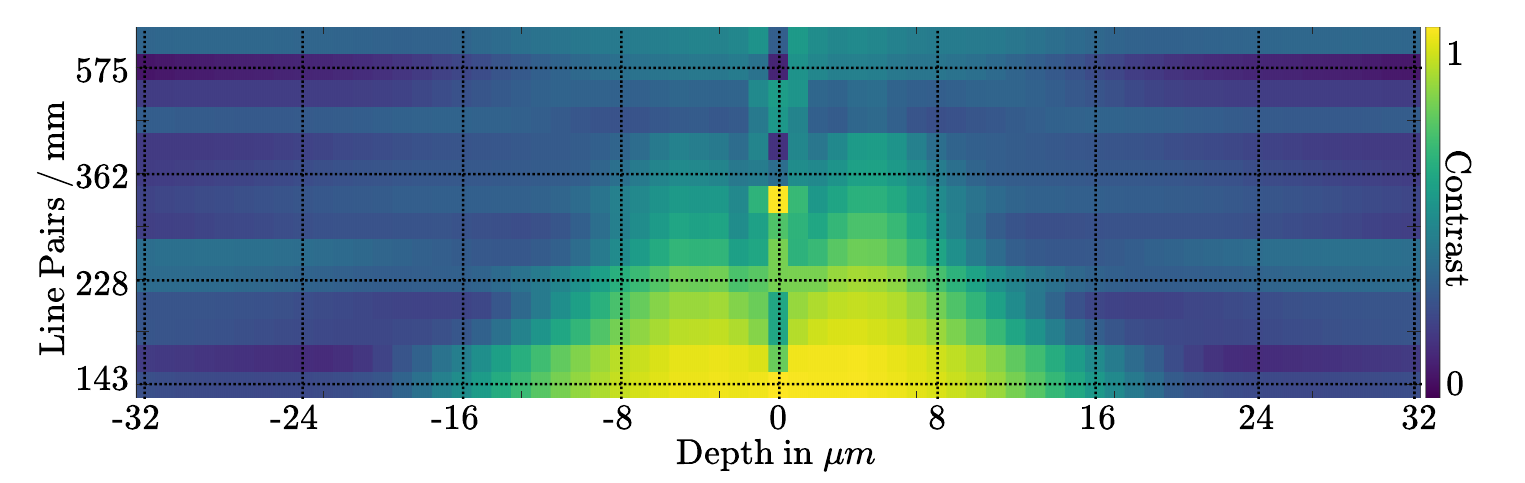}
        \caption{Simulated MTF}
    \end{subfigure}\\
    \begin{subfigure}[t]{\textwidth}
        \centering
        \includegraphics[width=\textwidth]{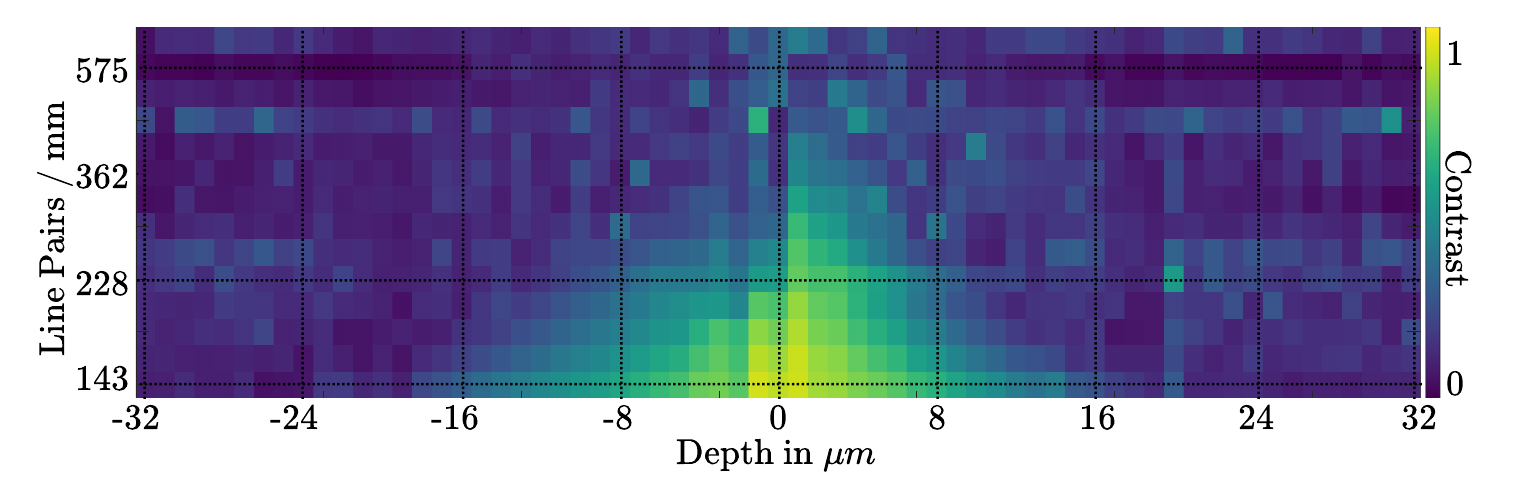}
        \caption{Measured MTF in our LFM}
    \end{subfigure}
    \caption{Comparison between (a) the simulated MTF and (b) our measured MTF.}
    \label{fig:MTF_compare}
\end{figure}

\section{Microscope MTF} \label{sec:A:realMTF}
In this section we measured the Modulation Transfer Function of our LMF as in section~\ref{fig:PerfSearch}.  Fig.~\ref{fig:MTF_compare} shows the comparison between an ideal MTF, obtained by using wave optics simulation, and the real MTF, measured from images of the USAF 1951 target. Both set of images were deconvolved with the aliasing-aware deconvolution algorithm \cite{ref:Stefanoiu:19}.

\section*{Acknowledgments}

We thank Daniel Sevilla Sánchez (SVI, the Netherlands), Sandra Frank and Stefan Tschanz (University of Bern, Switzerland) for installation and support of the HRM server and Eduard Babiychuk (University of Bern, Switzerland) for sharing the microscopy equipment. We are thankful to Gaby Enzmann, Josephine Mapunda and Elisa Kaba (University of Bern, Switzerland) for their help on sample preparation. We moreover acknowledge the Graduate School for Cellular and Biomedical Sciences (GCB) and the PhD program "Cutting Edge Microscopy" of the University of Bern. Microscopy was performed on equipment supported by the Microscopy Imaging Center (MIC), University of Bern, Switzerland.

\bibliography{bibliography}
\end{document}